\documentclass{aastex631}
\shorttitle{LASO}
\shortauthors{Cohen et al.}
\graphicspath{{./}{figures/}}

\begin{document}

\title{Longitudinally asymmetric stratospheric oscillation on a tidally locked exoplanet}

\author[0000-0001-5014-4174]{Maureen Cohen}
\affiliation{School of GeoSciences,
The University of Edinburgh \\
Edinburgh, EH9 3FF, UK}
\affiliation{Centre for Exoplanet Science, The University of Edinburgh, UK}

\author[0000-0001-7509-7650]{Massimo A. Bollasina}
\affiliation{School of GeoSciences,
The University of Edinburgh \\
Edinburgh, EH9 3FF, UK}

\author[0000-0002-1487-0969]{Paul I. Palmer}
\affiliation{School of GeoSciences,
The University of Edinburgh \\
Edinburgh, EH9 3FF, UK}
\affiliation{Centre for Exoplanet Science, The University of Edinburgh, UK}

\author[0000-0001-8832-5288]{Denis E. Sergeev}
\affiliation{Department of Mathematics \\
College of Engineering, Mathematics, and Physical Sciences, University of Exeter \\
Exeter, EX4 4QF, UK}

\author[0000-0002-1485-4475]{Ian A. Boutle}
\affiliation{Department of Astrophysics \\
College of Engineering, Mathematics, and Physical Sciences, University of Exeter \\
Exeter, EX4 4QF, UK}
\affiliation{Met Office, Fitzroy Road, Exeter EX1 3PB, UK}

\author[0000-0001-6707-4563]{Nathan J. Mayne}
\affiliation{Department of Astrophysics \\
College of Engineering, Mathematics, and Physical Sciences, University of Exeter \\
Exeter, EX4 4QF, UK}

\author[0000-0003-4402-6811]{James Manners}
\affiliation{Met Office, Fitzroy Road, Exeter EX1 3PB, UK}
\affiliation{Global Systems Institute, \\
University of Exeter \\
Exeter, EX4 4QF, UK}

\correspondingauthor{Maureen Cohen}
\email{s1144983@ed.ac.uk}



\begin{abstract}

Using a three-dimensional general circulation model, we show that the
atmospheric dynamics on a tidally locked Earth-like exoplanet, simulated with the planetary and orbital parameters of Proxima Centauri b, support
a longitudinally asymmetric stratospheric wind oscillation (LASO),
analogous to Earth's quasi-biennial oscillation (QBO).
In our simulations, the LASO has a vertical extent of 35--55~km, a period of 5--6.5
months, and a peak-to-peak wind speed amplitude of -70 to +130~ms$^{-1}$ with a
maximum at an altitude of 41~km.
Unlike the QBO, the LASO displays longitudinal asymmetries related to
the asymmetric thermal forcing of the planet and to interactions with the
resulting stationary Rossby waves.  The equatorial gravity wave
sources driving the LASO are localised in the deep convection region
at the substellar point and in a jet exit region near the western
terminator, unlike the QBO, for which these sources are distributed
uniformly around the planet.
Longitudinally, the western terminator experiences the highest wind
speeds and undergoes reversals earlier than other longitudes. The
antistellar point only experiences a weak oscillation with a very
brief, low-speed westward phase.
The QBO on Earth is associated with fluctuations in the abundances of
water vapour and trace gases such as ozone which are also likely to
occur on exoplanets if these gases are present. Strong fluctuations in
temperature and the abundances of atmospheric species at the
terminators will need to be considered when interpreting atmospheric
observations of tidally locked exoplanets.  

\end{abstract}

\keywords{Exoplanets (498) --- Exoplanet atmospheres (487)}


\section{Introduction} \label{sec:intro}

Rocky planets orbiting M-class stars have become a prime target for exoplanet research due to their prevalence and potential habitability \citep{yang_occurrence_2020, schwieterman_exoplanet_2018, kopparapu_revised_2013, heath_habitability_1999, scalo_m_2007, wandel_biohabitability_2018}. As the habitable zones (HZ) of cool M stars are found close to the host star, planets within the HZ have a high chance of being tidally locked \citep{barnes_tidal_2017, pierrehumbert_atmospheric_2019}. These planets receive stellar irradiation on only one hemisphere, an orbital configuration not found for planets in the solar system, which has a substantial impact on the planet's atmospheric dynamics and chemistry. 

An accurate understanding of atmospheric dynamics is important both to assessing the theoretical habitability of tidally locked rocky planets and interpreting observations. Theoretical studies of tidally locked rocky exoplanets have used 3-D atmospheric general circulation models (GCMs) to explore their possible atmospheric dynamics and climate states \citep{koll_temperature_2016, boutle_exploring_2017, wolf_assessing_2017, komacek_atmospheric_2019}. A key finding of GCM studies is that tidally locked planets within the HZ of M-class stars can retain temperate climates in spite of their uneven thermal forcing as a result of winds redistributing heat from their hot daysides to their cold nightsides \citep{merlis_atmospheric_2010, pierrehumbert_atmospheric_2019, wordsworth_atmospheric_2015, joshi_simulations_1997}. This day-night temperature gradient of a tidally locked exoplanet has already been observed in some cases, even for a rocky planet \citep{demory_map_2016}. 
However, most past studies have examined mean climate states, masking time-dependent phenomena which may impact observations by upcoming instruments such as the James Webb Space Telescope's Near Infrared Spectrograph (NIRSpec) \citep{molliere_observing_2017, morley_observing_2017} and the ESA Atmospheric Remote-sensing Infrared Exoplanet Large-survey mission \citep{venot_better_2018, tinetti_chemical_2018}. As observations are performed in a narrow time window, modes of variability with long periods such as analogues of the Earth's quasi-biennial oscillation (QBO) may result in observed temperatures or abundances that differ by an order of magnitude or more from predictions based on mean GCM states.
Other studies have used GCMs, coupled with synthetic planetary spectrum generators, to simulate observations of transmission spectroscopy for transiting exoplanets \citep{suissa_dim_2020, arney_pale_2017, may_water_2021, boutle_mineral_2020, lines_exonephology_2018}, enhancing the value of these models as an aid to interpreting observational data \citep{brogi_rotation_2016, louden_spatially_2015}. However, gaps in our understanding of the physics driving atmospheric phenomena on tidally locked planets, partly driven by differences in model structure and parameterisations \citep{sergeev_atmospheric_2020, fauchez_trappist-1_2021, turbet_trappist-1_2021, sergeev_trappist-1_2021}, may cause misinterpretations of atmospheric observations. 

Global 3-D models of tidally locked planets exhibit an area of deep convection around the substellar point \citep{hammond_rotational_2021, yang_stabilizing_2013, kopparapu_inner_2016}, leading to permanent cloud cover and heavy precipitation in this region \citep{boutle_exploring_2017, sergeev_atmospheric_2020, labonte_sensitivity_2020, yang_stabilizing_2013}. The tropospheric and stratospheric circulation is characterised by stationary Rossby waves of zonal wavenumber 1 induced by the day-night thermal forcing \citep{merlis_atmospheric_2010, showman_equatorial_2011, hammond_rotational_2021}. These waves have been identified as a driving source of the superrotating equatorial jets that develop in tidally locked simulations of both Earth-like planets and hot Jupiters \citep{showman_matsuno-gill_2010, showman_equatorial_2011, tsai_three-dimensional_2014, debras_eigenvectors_2019, debras_acceleration_2020}. The jets are associated with two pairs of cyclones-anticyclones: one pair forms in the northern hemisphere, rotating clockwise on the dayside and anticlockwise on the nightside, while the other pair forms in the southern hemisphere and rotates in the opposite directions. The cyclones contribute a zonal wind vector component near the equator that points westward in the substellar region and eastward in the antistellar region. This results in a background flow which is more eastward on the nightside than on the dayside, a pattern that holds true in both the troposphere and the stratosphere. The day-night thermal forcing additionally creates a region of rising air on the dayside and subsiding air on the nightside \citep{hammond_rotational_2021, joshi_simulations_1997}. 

On Earth, the background flow determines the zonal structure of the QBO. In the QBO, the stratospheric zonal wind forms a pattern of vertically stacked jets flowing in opposite directions. The pattern propagates downwards from roughly $20-40$ km at a speed of $1$ km per month such that the wind direction at a given height reverses over time with a mean periodicity of $26-28$ months, with substantial variation from cycle to cycle \citep{dunkerton_middle_2015, braesicke_middle_2015}. The oscillation of the zonal wind is accompanied by oscillations in air temperature and trace gases. 
The basic mechanism driving the QBO was described by \cite{lindzen_theory_1968} and \cite{plumb_interaction_1977} and reproduced in a laboratory setting by \cite{plumb_instability_1978}. The minimum requirements for the QBO to occur are a background flow that is a function of altitude and at least two gravity waves propagating in opposite directions horizontally, as well as upwards vertically \citep{plumb_interaction_1977}. An eastward propagating gravity wave will travel upwards until it encounters an eastward perturbation in the background flow, where it is absorbed, accelerating the flow in the eastward direction. Westward propagating waves pass through this area of eastward flow until they reach a westward perturbation and are absorbed in turn. The resulting pattern of stacked jets propagates downwards over time as the bottom of each shear zone is accelerated and the lowest jet dissipates due to viscous diffusion \citep{plumb_interaction_1977, dunkerton_middle_2015}. Although equatorially trapped Rossby and Kelvin waves contribute some momentum flux, it is believed that up to $80\%$ of the momentum flux driving the QBO is provided by convectively generated gravity waves \citep{dunkerton_role_1997, lane_gravity_2015}. Gravity waves are generated when the forces of gravity or buoyancy react against vertical perturbations in a fluid medium. The waves travel internally within the fluid until they are absorbed at critical levels where the velocity of the background flow is comparable to the wave's horizontal phase speed \citep{booker_critical_1967}, making the planet's mean flow an important element in the phenomenon.

Here, we describe the first reported simulation of a stratospheric oscillation, analogous to the QBO on Earth, on a tidally locked planet with an Earth-like atmospheric composition. This new stratospheric oscillation consists of periodic reversals in the direction of the zonal wind, an associated cooling and warming of the equatorial atmosphere, and fluctuations in the stratospheric water vapour content. While analogous to the QBO in its causal mechanism, the oscillation displays longitudinal asymmetries arising directly from the planet's tidally locked state. Accordingly, we refer to the oscillation on a tidally locked planet as a longitudinally asymmetric stratospheric oscillation (LASO) and reserve the term QBO for the phenomenon on Earth. To put our results into context, throughout the paper we compare and contrast key features of the tidally locked LASO with the QBO on Earth \citep{baldwin_quasi-biennial_2001}. 

Section \ref{sec:methods} describes the global 3-D model we use to describe atmospheric dynamics on a tidally locked planet, the metrics we use to characterize the LASO, and the radiative transfer model we use to simulate JWST observations of the fluctuations in atmospheric water vapour content caused by the LASO. Section \ref{sec:results} describes the mechanism underpinning the LASO and characteristics of the LASO that differ from those of the QBO. We also consider the effect of LASO-related water vapour fluctuations on transit observations. In Section \ref{sec:discussion} we discuss the implications of the LASO for future spectral atmospheric observations and the sensitivity of the LASO to differences in sub-grid scale model parameterisations. We conclude the paper in Section \ref{sec:conclusion}.

\section{Methods} \label{sec:methods}

\subsection{Model description} \label{subsec:model}

We use the Global Atmosphere 7.0 configuration of the Met Office Unified Model to simulate the atmosphere of a tidally locked Earth-like exoplanet, nominally Proxima Centauri b \citep{walters_met_2019}. For brevity, we provide a basic description of the model and refer the reader to \cite{mayne_using_2014}, \cite{boutle_exploring_2017}, and \cite{sergeev_atmospheric_2020} for more details.

We run the model at a horizontal resolution of $2^\circ$ latitude by $2.5^\circ$ longitude, and with 60 vertical levels from the surface to up to 85~km (Table \ref{vertlevs}). The atmospheric levels are identical to those in \cite{boutle_exploring_2017} up to $38$ km. The remaining $47$ km are accounted for by an additional $22$ levels. 
Table \ref{parameters} shows the orbital and planetary parameters used in this study, following \cite{boutle_exploring_2017}, with data taken from \cite{anglada-escude_terrestrial_2016} and \cite{turbet_habitability_2016}. Proxima Centauri b is simulated as an aquaplanet with a slab ocean \citep{frierson_gray-radiation_2006}, with a heat capacity of $10^7$~{J K$^{-1}$ m$^{-2}$} representing a mixing layer of 2.4~m. We specify an atmospheric composition of 100 \% nitrogen with a trace fixed $\mathrm{CO_2}$ abundance of $5.941 \times 10^{-04} \mathrm{kg kg^{-1}}$. The model includes interactive water vapour and a full water cycle with unlimited evaporation from the slab ocean, precipitation in the liquid and ice phases, and a cloud scheme. The stellar spectrum for Proxima Centauri was taken from BT-Settl \citep{rajpurohit_effective_2013} with $T_{eff} = 3000$ K, g= $1000 \mathrm{m s^{-2}}$, and metallicity $=0.3$ dex. The star is treated as quiescent.
We spin up the model for $900$ Earth days from an equilibrium state taken from a previous run of the Proxima Centauri b model. The simulation was determined to be in equilibrium when the incoming and outgoing radiation at top-of-atmosphere were balanced, evaporation balanced precipitation in the global mean, and the surface temperature no longer showed a long-term upward or downward trend. The experiments we report were run for $1800$ Earth days in total, and we analysed a sampling period of $900$ days after equilibrium, which is long enough to include multiple oscillations.
We define the substellar point to be $0^{\circ}$ longitude and latitude. The antistellar point is located at $180^{\circ}$ longitude and the eastern and western terminators at $90^{\circ}$E and $90^{\circ}$W, respectively. Hereafter, days refer to Earth days.

\begin{table}
\centering
\begin{tabular}{llllcllllll}
\hline \hline
Level & 1 & 2 & 3 & 4 & 5 & 6 & 7 & 8 & 9 & 10 \\
\hline 
Height (m) & 10.0 & 49.9 & 130.0 & 249.99 & 410.0 & 610.0 & 849.9 & 1129.9 & 1449.9 & 1810.0 \\
\hline \hline
Level & 11 & 12 & 13 & 14 & 15 & 16 & 17 & 18 & 19 & 20 \\
\hline 
Height (m) & 2209.9 & 2650.0 & 3129.9 & 3650.0 & 4210.0 & 4810.0 & 5449.9 & 6129.9 & 6849.9 & 7609.9 \\
\hline \hline
Level & 21 & 22 & 23 & 24 & 25 & 26 & 27 & 28 & 29 & 30 \\
\hline 
Height (m) & 8409.9 & 9249.9 & 10130.0 & 11050.0 & 12001.0 & 13001.0 & 14050.0 & 15130.0 & 16250.0 & 17410.0 \\
\hline \hline
Level & 31 & 32 & 33 & 34 & 35 & 36 & 37 & 38 & 39 & 40 \\
\hline 
Height (m) & 18590.0 & 19770.1 & 20950.3 & 22131.3 & 23313.9 & 24499.5 & 25690.3 & 26889.2 & 28100.2 & 29328.5 \\
\hline \hline
Level & 41 & 42 & 43 & 44 & 45 & 46 & 47 & 48 & 49 & 50 \\
\hline 
Height (m) & 30580.3 & 31863.4 & 33186.9 & 34562.1 & 36001.5 & 37520.0 & 39134.3 & 40863.5 & 42729.0 & 44754.9 \\
\hline \hline
Level & 51 & 52 & 53 & 54 & 55 & 56 & 57 & 58 & 59 & 60 \\
\hline 
Height (m) & 46967.7 & 49396.9 & 52074.8 & 55036.7 & 58321.5 & 61971.1 & 66030.9 & 70550.2 & 75581.7 & 81182.4 \\
\hline \hline
\end{tabular}
\caption{Vertical levels and heights (layer centres)  used in the Unified Model to describe the atmospheric dynamics of an Earth-like tidally locked Proxima Centauri b.}
\label{vertlevs}
\end{table}

\begin{table}
\centering
\begin{tabular}{llllcllll}
\hline
Parameter            & Value  \\
\hline \hline
Semi-major axis (AU) & 0.0485 \\
Stellar irradiance (W m$^{-2}$) & 881.7 \\
Orbital period (Earth days) & 11.186 \\
Rotation speed (rad s$^{-1}$) &  6.501$\times$10$^{-6}$      \\
Eccentricity ($\cdot$) & 0     \\
Obliquity ($\cdot$) & 0   \\
Radius (km) & 7160   \\
Acceleration due to gravity (m s$^{-2}$) & 10.9 \\
\hline
\end{tabular}
\caption{Simulation parameters used to describe an Earth-like tidally locked Proxima Centauri b.}
\label{parameters}
\end{table}

Earth GCMs typically require both resolved and parameterised waves to generate a QBO \citep{garfinkel_qbo_2021}. Past studies have shown that the Unified Model can reproduce a realistic oscillation \citep{scaife_realistic_2000, scaife_impact_2002}. Our model can sustain resolved long wavelength gravity waves and inertio-gravity waves generated by physical and numerical imbalances in the flow (as discussed in Section \ref{sec:results}). The model also includes a sub-grid scale non-orographic gravity wave parametrisation \citep{warner_propagation_1996, warner_ultrasimple_2001, bushell_parameterized_2015}. Within this parameterisation, an unsaturated spectrum of gravity waves is launched near the surface and carries a vertical flux of horizontal wave pseudo-momentum upwards in equal amounts towards the north, south, east, and west. Upward-propagating flux is diminished by Doppler modification and density amplification. Wave pseudo-momentum is conserved by an opaque model lid which sets the outgoing flux to zero \citep{bushell_umdp_2021}. The scheme includes a scale factor for the vertical flux and a characteristic (spectrum peak) wavelength which must be tuned to the vertical resolution of the model to achieve a QBO. 
We use a factor enhancement for the vertical flux of horizontal pseudo-momentum of 1.2 and a characteristic wavelength of 4.3~km with model level heights shown in Table \ref{vertlevs}. These are the standard values used in the GA 7.0 for Earth with the model levels in Table \ref{vertlevs} and we retain them in order to simulate Proxima Centauri~b as an Earth-like planet. The gravity wave scheme can be run with two different gravity wave source locations: 1) a standard scheme in which the gravity wave launch flux is globally invariant, distributed uniformly around the model; and 2) an experimental scheme in which the gravity wave launch flux is determined by local precipitation, developed based on an empirical correlation found between gravity waves and precipitation on Earth \citep{bushell_parameterized_2015}. Our study is based on the standard globally invariant scheme. To test whether our results are robust against the choice of scheme, we ran the experimental precipitation-dependent scheme and found that the results are comparable with the exception of the amplitude of the longitudinal asymmetry of the mean flow, which we discuss in Section \ref{sec:results}.

In addition to the control experiment described below, we performed a sensitivity run to understand the influence of horizontal resolution by increasing it to $2^\circ$ latitude by $2^\circ$ longitude. We performed a separate sensitivity test by decreasing the timestep from 20 to 2 minutes. We found that these sensitivity runs did not result in major changes from the control experiment.

\subsection{Metrics used to characterise the LASO} \label{sec:metrics}

\subsubsection*{Gravity wave-induced acceleration}

To confirm that the reversal in the direction of the stratospheric jets is caused by absorption of gravity waves, we calculate the acceleration of the atmosphere in each model gridbox due to gravity waves, $F$, as: 
\begin{equation}
    F = \frac{\partial{(u^{\prime}w^{\prime})}}{\partial{z}},
    \label{eqn:F}
\end{equation}
where $u^{\prime}$ is the high-pass filtered zonal anomaly of the zonal wind, $w^{\prime}$ is the time anomaly of the vertical wind, and \emph{z} is the height. This definition is adapted from the formulation of the wave momentum flux described by \cite{plumb_interaction_1977} and used in its original form by \cite{showman_atmospheric_2019} to identify shear transition zones between alternating jets.

We modified $u^{\prime}$ to better describe the characteristic flow on a tidally locked planet. The zonal anomaly is usually defined as the deviation of that quantity from its zonal mean and, in a zonally symmetric baseline flow, identifies transient phenomena. However, tidally locked planets lack zonal symmetry. The planet's stationary Rossby waves skew the zonal anomaly so that it represents the Rossby waves rather than transient flows. To adapt $u^{\prime}$ to this asymmetrical background flow, we use a high-pass filter to remove the influence of waves with zonal wavenumber less than 5 (i.e., long Rossby waves), following \cite{sugimoto_generation_2021}. The remaining waves are resolved long wavelength gravity and inertio-gravity waves. 
Additionally, the zonal anomaly of the vertical wind \emph{w} would be impractical for identifying waves because rising air on the dayside and subsiding air on the nightside cancel each other out when a zonal mean is calculated. We instead use the time anomaly to distinguish transient, wave-like fluctuations in \emph{w} from the background. The resulting metric identifies wave-induced acceleration caused by resolved gravity and inertio-gravity waves only.

\subsubsection*{Latitudinal extent}

We expect that a QBO-like oscillation on an exoplanet may extend to higher latitudes than Earth's QBO, covering a larger portion of the terminators, which may affect transit spectroscopy. To characterise the latitudinal extent $L$ of the LASO, we use a formula derived by \cite{haynes_latitudinal_1998} for the latitudinal extent of the QBO:
\begin{equation}
    L = (\frac{\sigma}{\alpha})^{1/4}(\frac{ND}{\beta_0})^{1/2},
    \label{eqn:L}
\end{equation}
where $\sigma$ is the frequency of time variation of the applied force, $\alpha$ is the radiative damping rate, \emph{N} is the buoyancy frequency, \emph{D} is the depth scale of the force, and $\beta_0$ is the meridional gradient of the Coriolis parameter evaluated at the equator. We assume that all quantities except $\beta_0$ are identical to those for Earth, and focus our analysis on the impact of differences in the Coriolis factor.

\subsubsection*{Period}

The period of the LASO is of interest because the duration of the temperature and atmospheric composition oscillations may affect  observations and in particular could cause discrepancies between repeated observations of the same target exoplanet. We determine the period $T$ of the LASO using a wavelet analysis of the wind velocity time series at the altitude where the LASO has its greatest amplitude. 
Following \cite{plumb_interaction_1977}, the period of the oscillation is inversely proportional to the momentum flux:
\begin{equation}
    T = \eta \frac{\hat{k} \hat{c}^3}{\hat{N} \hat{\mu} \hat{F}}, 
    \label{eqn:T}
\end{equation}
where $\hat{k}$, $\hat{c}$, $\hat{N}$,  $\hat{\mu}$, and $\hat{F}$ are scaling factors for the wavenumber and phase speed of the wave driving the oscillation, the buoyancy frequency, the thermal dissipation rate, and the momentum flux, respectively. The parameter $\eta$ is a dimensionless number that depends on the specific details of the problem. In Plumb's simple numerical model, the wavenumber and phase speed of the waves can be specified, while in the real atmosphere the oscillation is driven by a broad spectrum of waves whose constituents vary with location and time. However, the inverse relationship between the period and the momentum flux at the lower boundary of the QBO shear zone is also accepted to exist for the real QBO \citep{dunkerton_middle_2015}. We assume all other values are identical to those for Earth and focus on the impact of the wave momentum flux $\hat{F}$ on the period.

\subsubsection*{Lagrangian Rossby number}

To understand how a tidally locked orbital configuration affects where gravity waves are generated on the planet we use the Lagrangian Rossby number, $Ro^{(L)}$, to identify resolved gravity wave sources. $Ro^{(L)}$ is not an exclusive measure of gravity wave sources, but is often used to detect unbalanced flows which give rise to gravity waves \citep{zhang_survey_2000, lin_mesoscale_2007, sugimoto_generation_2021}. It is defined as the ratio of the acceleration of a parcel of air to the Coriolis acceleration and essentially measures the local departure from geostrophy:
\begin{equation}
    Ro^{(L)} = \frac{\mid \frac{\partial{v_h}}{\partial{t}}+ v_h \cdot
      \nabla v_h \mid}{f \mid v_h \mid},
    \label{eqn:rol}
\end{equation}
where $v_{h}$ is the horizontal wind vector and \emph{f} is the Coriolis parameter. 

We modify $Ro^{(L)}$ in several respects to adapt the metric to a slowly rotating planet. Following \cite{sugimoto_generation_2021}, we disregard the local wind tendency term, $\frac{\partial{v_h}}{\partial{t}}$, and fix the value of \emph{f} near the equator ($\leq \pm 10^{\circ}$) equal to \emph{f} at $10^{\circ}$ to extend the metric to cover the equatorial region. As the slowly rotating Proxima Centauri b is not in geostrophic balance, the resulting values of $Ro^{(L)}$ are very large. We normalise $Ro^{(L)}$ by dividing by the minimum value, which we consider the baseline ageostrophy for our simulation. A larger value indicates a relative local increase in imbalances in the flow. We also exclude the top five model levels and latitudes above $\pm 40^{\circ}$ to avoid interference from instability at the model top and poles.

\subsection{NASA Planetary Spectrum Generator}

To describe the spectrum of the atmospheric water vapour content that might be observed by the James Webb Space Telescope's NIRSpec instrument, we use the NASA Planetary Spectrum Generator \citep{villanueva_planetary_2018}, publicly available at \url{https://psg.gsfc.nasa.gov/}. The NIRSpec instrument has a range of 0.6 to 5.3 $\mathrm{\mu m}$, covering water vapour features in the infrared at 1.4, 1.8, and 2.7 $\mathrm{\mu m}$, with spectral resolutions (R) of 100, 1000, and 2700. As inputs to the PSG, we use the basic orbital, planetary, and stellar parameters for the Proxima Centauri system, as well as pressure, temperature, altitude, H$_2$O, N$_2$, and CO$_2$ data from our UM simulations. We generate a transit spectrum for each latitude at both the eastern and western terminators (i.e., every $2^\circ$ of latitude, for a total of 180 individual spectra) and average the outputs to get a final spectrum, following \cite{suissa_dim_2020}.

\section{Results} \label{sec:results}

\subsection{Longitudinally asymmetric stratospheric oscillation} \label{subsec:laso}

\begin{figure}

\gridline{\fig{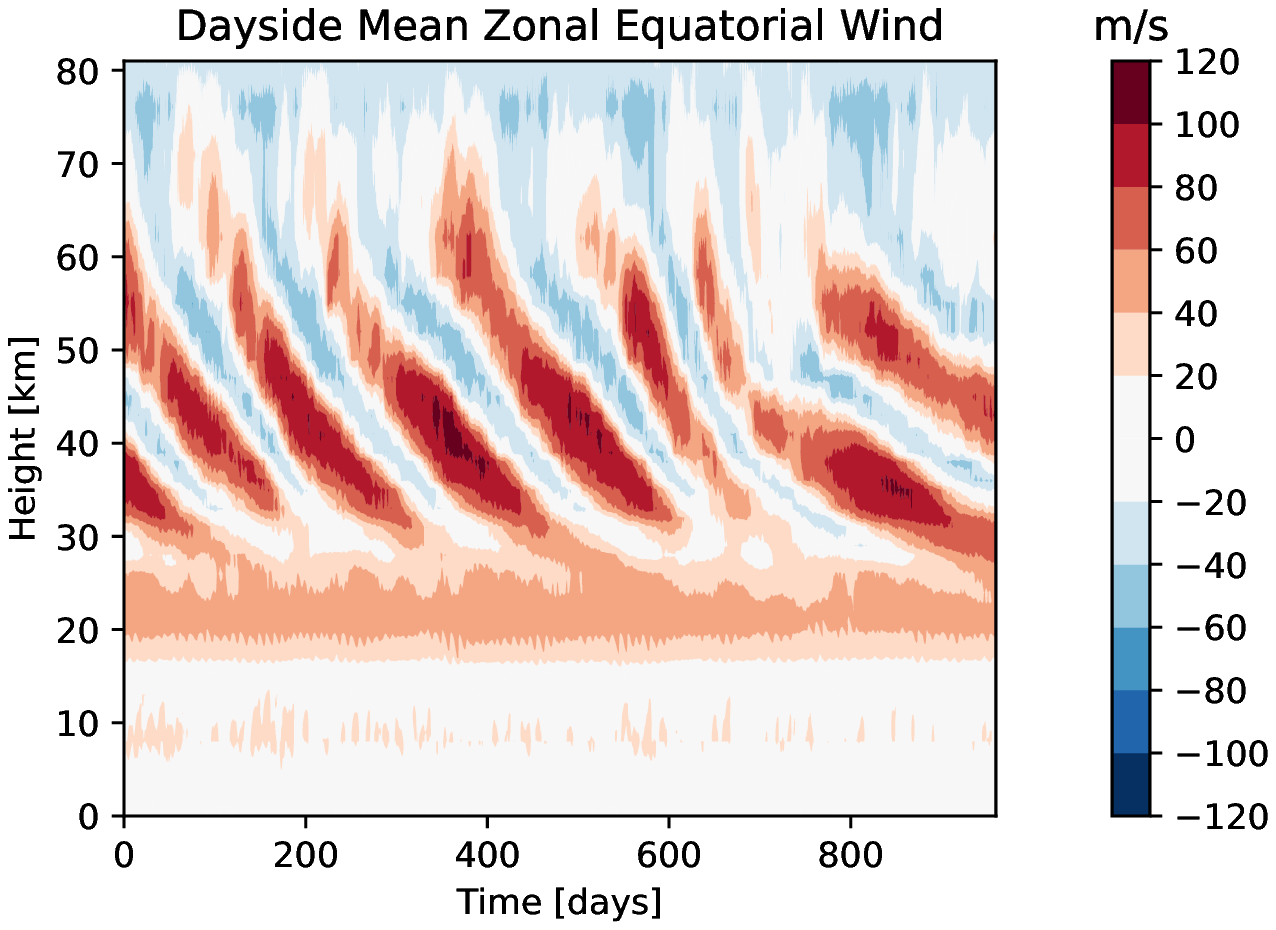}{0.5\textwidth}{(a) Dayside zonal mean zonal wind} \fig{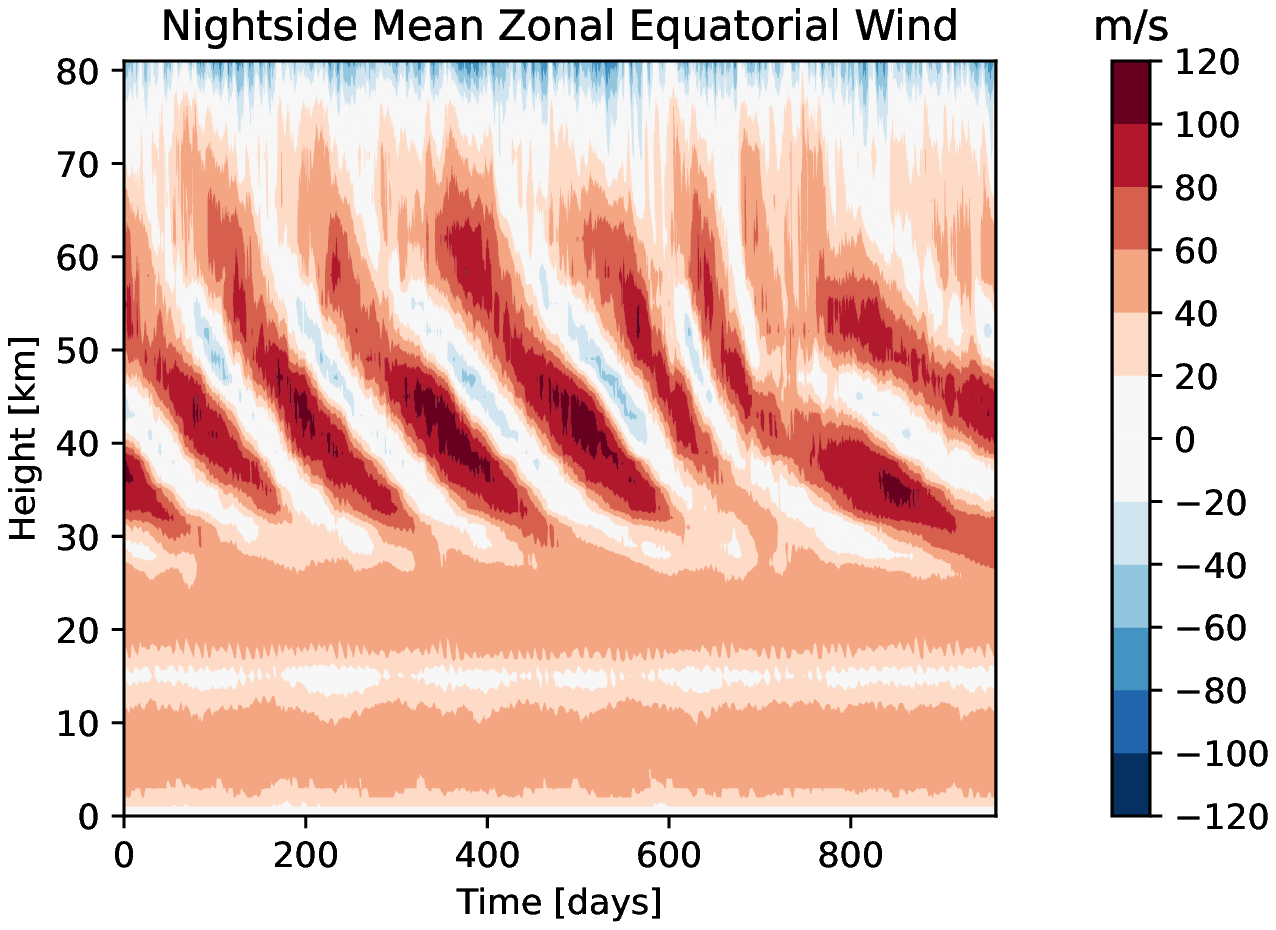}{0.5\textwidth}{(b) Nightside zonal mean zonal wind}}

\gridline{\fig{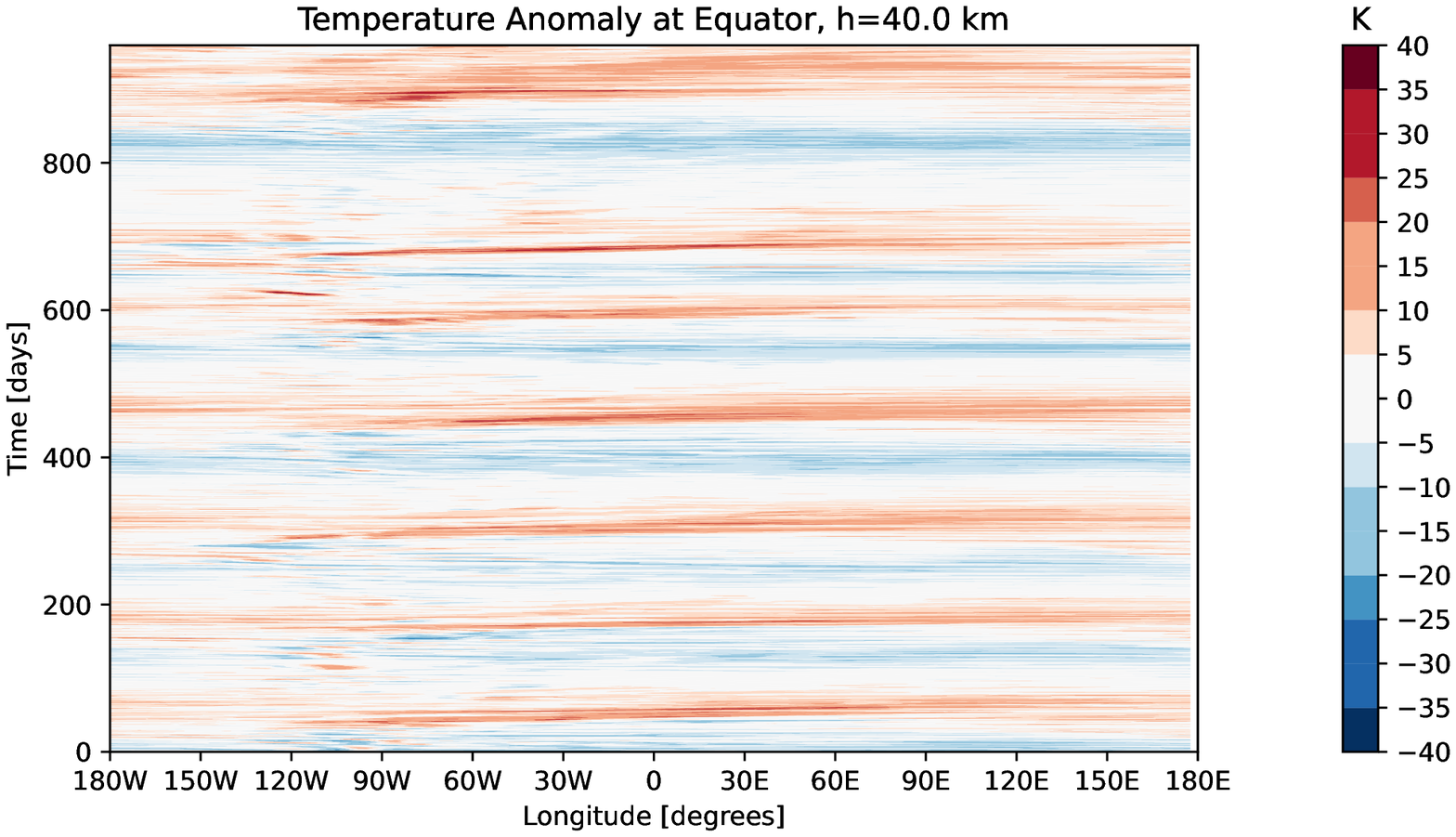}{0.9\textwidth}{(c) Temperature anomaly at equator at 40 km}}
\caption{Hovmöller diagrams of a) dayside and b) nightside zonal mean equatorial winds on the simulated planet and c) the absolute temperature anomaly at 40 km. The data are sampled every six hours.}
\label{fig:laso}
\end{figure}

Figure \ref{fig:laso} shows the wind speed and temperature oscillations in our simulation of the atmosphere of the tidally locked Proxima Centauri~b. The period of the LASO is much shorter ($5-6.5$ months) than that of the QBO on Earth ($26-28$ months). There is also an asymmetry between the dayside and nightside. Nightside zonal winds are more positive (eastward) than dayside winds and have shorter westward phases. Consistent with the theory that describes the QBO, the dayside-nightside asymmetries in the mean planetary-scale circulation, in particular the dayside-nightside differences in wind speeds, are the primary factor responsible for the asymmetry of the LASO. We find that the location of gravity wave sources also affects the local flow. Figure \ref{fig:laso}c shows temperature fluctuations with a peak-to-peak amplitude of $\sim 60-70$ K, more than an order of magnitude larger than the variation of $\sim 4$ K associated with the QBO \citep{dunkerton_middle_2015}.

\begin{figure}

\gridline{\fig{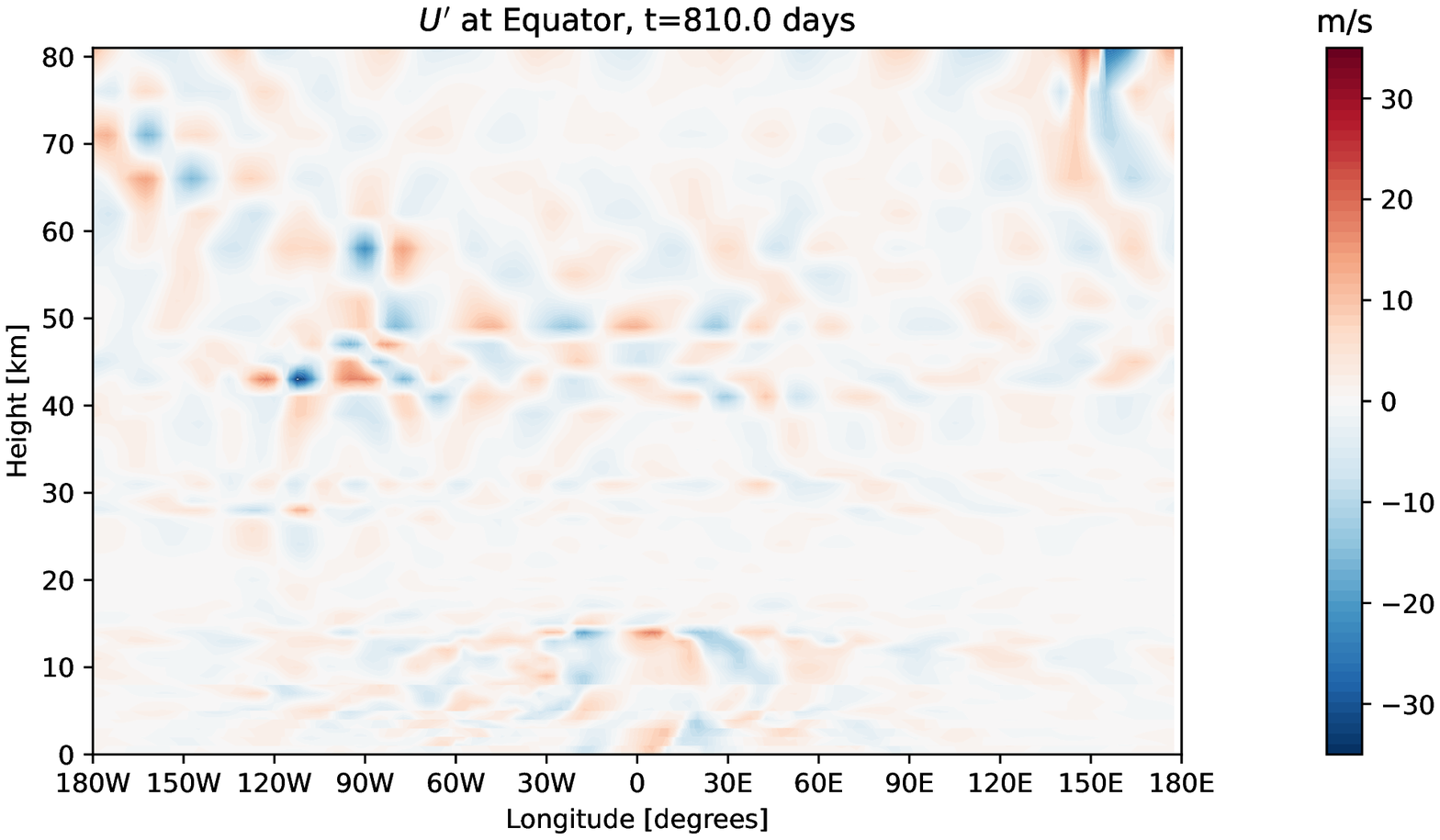}{1.0\textwidth}{(a) The zonal anomaly of the zonal wind as a function of altitude, with waves of wavenumber less than 5 filtered out}}
\gridline{\fig{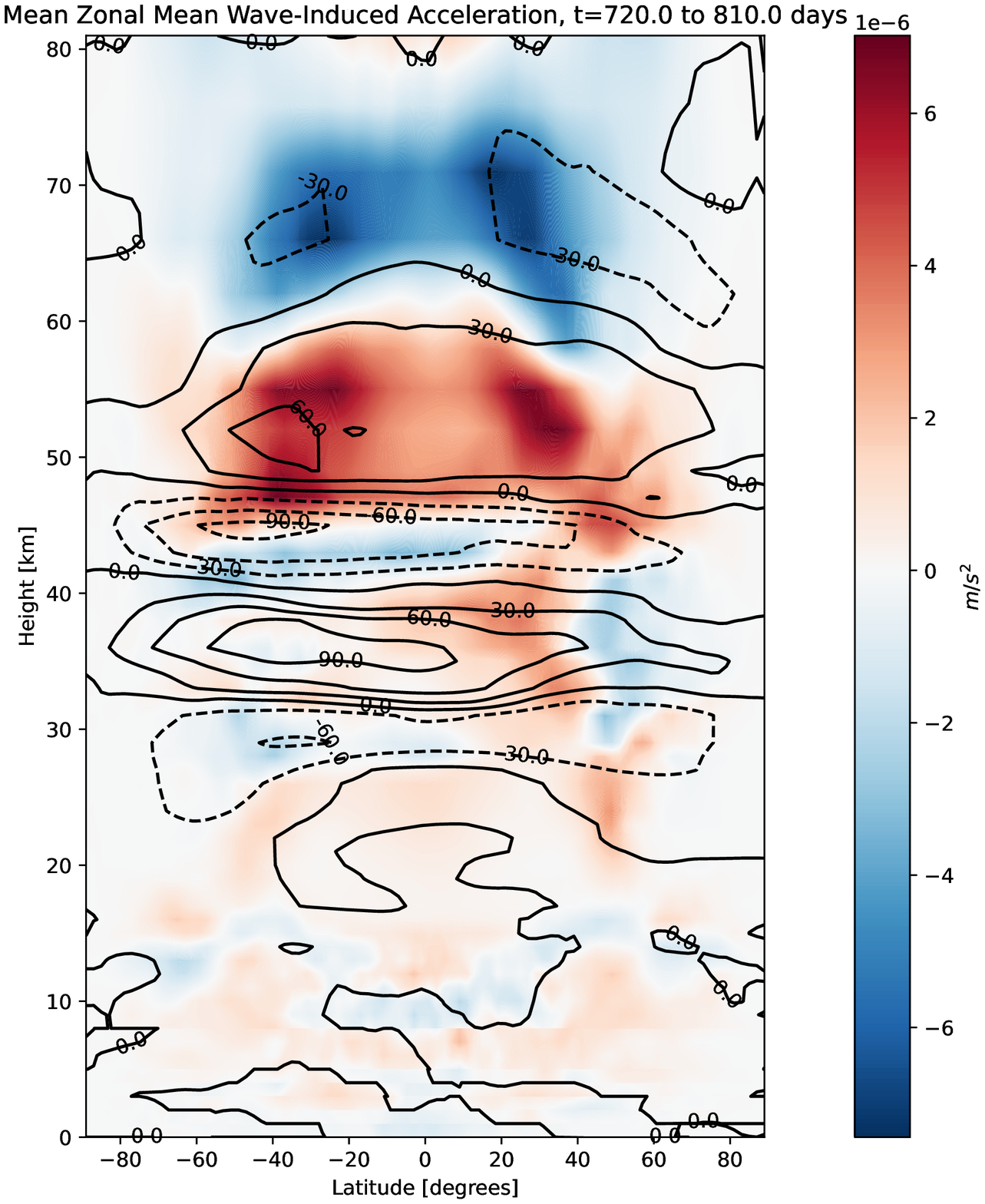}{0.5\textwidth}{(b) The zonal mean gravity wave-induced acceleration, averaged over time and compared to the change in zonal mean zonal wind direction}}

\caption{a) Zonal wind anomalies at the equator as a function of altitude, with zonal wavenumbers $<$5 removed, corresponding to a six-hour snapshot from simulation day 810 from the simulation sampling period. b) Zonal mean wave-induced acceleration due to gravity waves, overlaid with the change in zonal mean wind speeds. The values represent a 90-day mean from days 720 to 810 of the simulation sampling period.}
\label{fig:waveacc}
\end{figure}

To confirm that the LASO is generated by the same physical mechanism as the QBO, we calculated the wave-induced acceleration due to gravity waves and compared it to the change in zonal wind direction over time. Where the oscillation is driven by gravity waves, the direction of the acceleration caused by gravity waves should match the direction of the change in the zonal wind throughout the vertical extent of the LASO. Figure \ref{fig:waveacc}a shows the filtered $u^{\prime}$ field with zonal wavenumbers smaller than 5 removed (Section \ref{sec:metrics}). We find positive and negative fluctuations in wind speeds due to resolved long wavelength gravity waves, with shear zones at 30~km, 41~km, and 50~km that correspond to regions where the flow changes direction. Waves with a leftward tilt correspond to westward-propagating modes and those with a a rightward tilt correspond to eastward-propagating modes. The band of reduced $u^{\prime}$ intensity between $15$ km and $30$ km corresponds to the permanent superrotating equatorial jet in the troposphere. Figure \ref{fig:waveacc}b shows the zonal mean gravity wave-induced acceleration calculated using Equation \ref{eqn:F} and averaged over 90 days, corresponding to approximately half of a LASO cycle. The contour lines show the change in zonal wind speed over the same 90 days. Areas of positive (negative) wind direction change match well with areas of gravity wave-induced eastward (westward) acceleration. Regions below 25$-$30~km show little change in zonal wind speeds and negligible gravity wave-induced acceleration.

\subsection{Latitudinal extent of LASO}\label{subsec:latitudinal}

\begin{figure}
    \plotone{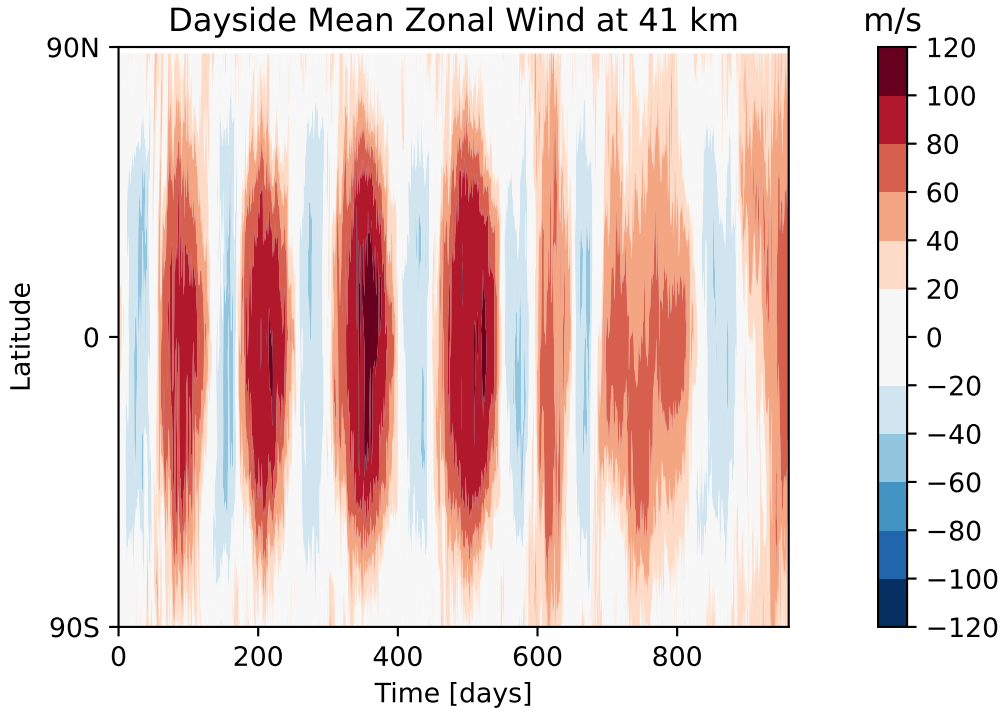}
    \caption{Latitudinal extent of the LASO, described by simulation data at a resolution of six hours.}
    \label{fig:topview}
\end{figure}

The LASO extends to higher latitudes than the QBO on Earth. The latitudinal extent of the QBO, defined here by the full width at half maximum of a Gaussian fit to the horizontal cross section of the zonal wind at the height of the QBO maximum \citep{schenzinger_defining_2017}, is roughly 15$^{\circ}$S-15$^{\circ}N$ \citep{braesicke_middle_2015}. Figure \ref{fig:topview} shows that the LASO extends from about 60$^{\circ}$S--60$^{\circ}$N by this definition. As the LASO on a slowly rotating tidally locked planet extends to higher latitudes, the oscillations in air temperature and in abundances of atmospheric species will cover more of the terminators and potentially have a larger impact on observations.

Previous work suggested that at higher latitudes the Coriolis force would counteract the acceleration due to gravity waves, limiting the QBO to equatorial regions on the Earth \citep{lindzen_theory_1968}. Equation \ref{eqn:L} gives a value for Earth of about $L_{\rm Earth}$=10$^{\circ}$ \citep{baldwin_quasi-biennial_2001}, which underestimates the actual extent of 15$^{\circ}$ \citep{dunkerton_climatology_1985} by $\simeq$ 33\%. The Coriolis force on Proxima Centauri b is an order of magnitude weaker than on Earth due to the planet's slower rotational period of 11.2~days ($\Omega$=6.5$\times$10$^{-6}$ rad/s). If we assume the values of the other quantities in Eqn. \ref{eqn:L} for Proxima Centauri b are identical to those for Earth, this gives a latitudinal extent of $\sqrt{10}L_{Earth}$ or $\simeq$47$^{\circ}$. Scaling this value to account for the 33\% underestimation gives the latitudinal range simulated by the model. The broader extent of the LASO compared to the equatorial QBO agrees with the description of slowly rotating tidally locked planets as `all tropics' \citep{showman_atmospheric_2013}. A LASO on a slow rotator may therefore have a larger impact on transit spectroscopy than one on a more rapidly rotating planet. 

\subsection{Period of LASO}\label{subsec:period}

Another major difference between the LASO and the QBO is the period of oscillation. The period of the QBO has a large variability, with a mean value of 26--28 months \citep{braesicke_middle_2015}. Our calculations show that the LASO in our simulation of Proxima Centauri b's atmosphere has a period of 5--6.5 months. The period of the QBO is not directly related to the planet's rotation rate (\cite{plumb_instability_1978} and Eqn. \ref{eqn:T}). Because a tidally locked planet does not rotate with respect to its star, the oscillation is also not related to seasonal and annual variations. The shorter period of LASO instead implies stronger wave momentum flux and wave-induced acceleration.

\begin{figure}
    \plotone{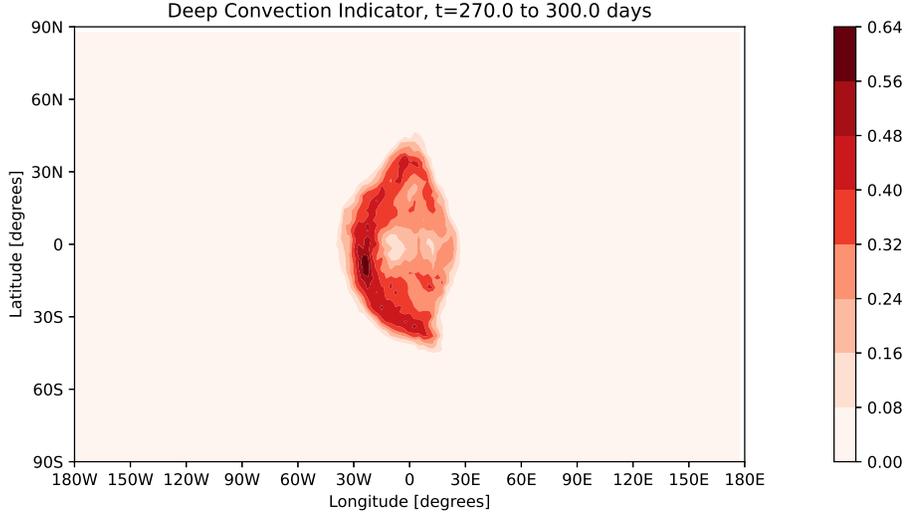}
    \caption{Deep convection indicator, averaged over a 30-day period (1: convection is present in the gridbox; 0: not present).
    In the Unified Model, convection is diagnosed by the surface buoyancy flux and an undilute parcel ascent from grid points where this flux is positive. It is then categorised as deep if the parcel reaches neutral buoyancy above $2.5\mathrm{km}$ or the freezing level (whichever is higher), and if the vertical velocity is greater than $0.02 \mathrm{ms^{-1}}$ within a layer at least $1.5 \mathrm{km}$ thick above the level of neutral buoyancy \citep{walters_met_2019}.}
    \label{fig:deepconvection}
\end{figure}

On Earth, up to $80\%$ of the QBO forcing originates from convectively generated gravity waves \citep{lane_gravity_2015}. Figure \ref{fig:deepconvection} shows that deep convection in our simulation of Proxima Centauri b occurs within an area of $30^\circ$ of the substellar point. This area has the shape of a pointed oval and is asymmetrical, with convection consistently more intense at the western edge of the region. This region of deep convection is collocated with a gravity wave source (as shown in Figure \ref{fig:lrn}b). The accepted dependence of the QBO's period on the gravity wave momentum flux as expressed in Equation \ref{eqn:T} supports the conclusion that greater wave momentum flux, driven by the intense convection, accounts for the shorter period of the LASO compared to the QBO on Earth. Accordingly, we find higher wind speeds in the LASO than the QBO. The QBO has maximum eastward winds of 15~ms$^{-1}$ and maximum westward winds of -30~ms$^{-1}$ \citep{dunkerton_middle_2015}, while the LASO's eastward wind speeds reach 100~ms$^{-1}$ and westward winds reach -75~ms$^{-1}$.

\subsection{Longitudinal asymmetry of the LASO}\label{subsec:asymmetry}

The LASO exhibits longitudinal asymmetries that do not have an analogy on Earth. We show below that this is related to underlying asymmetries in the steady-state atmospheric dynamics of tidally locked planets and ultimately to the asymmetrical thermal forcing. 

As described above, the emergence of the QBO is associated with gravity wave absorption at critical layers in the atmosphere where the speed of the background flow is similar to the phase speed of the wave, leading to wave breaking and the deposition of momentum. Longitudes and heights where the background flow is faster will absorb faster gravity waves and experience stronger acceleration, perpetuating the asymmetry. This accounts for the stronger nightside eastward winds seen in Figure \ref{fig:laso}b, as the nightside experiences higher eastward winds due to the Rossby wave contribution to the zonal wind, as well as for the height at which the LASO has its maximum amplitude (41~km), which is also where the Rossby waves are most intense.

\begin{figure}

\gridline{\fig{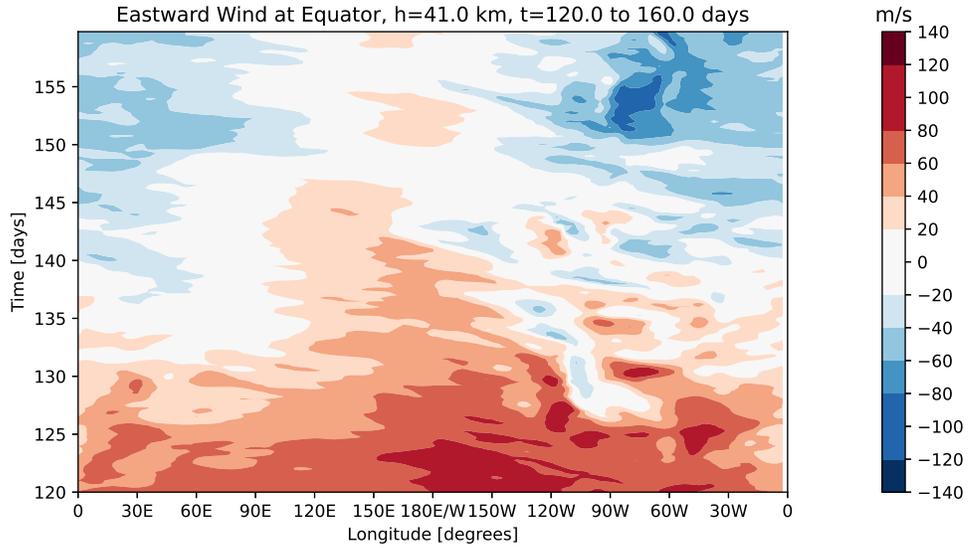}{0.9\textwidth}{(a) Transition from eastward to westward phase at 41 km}}
    \fig{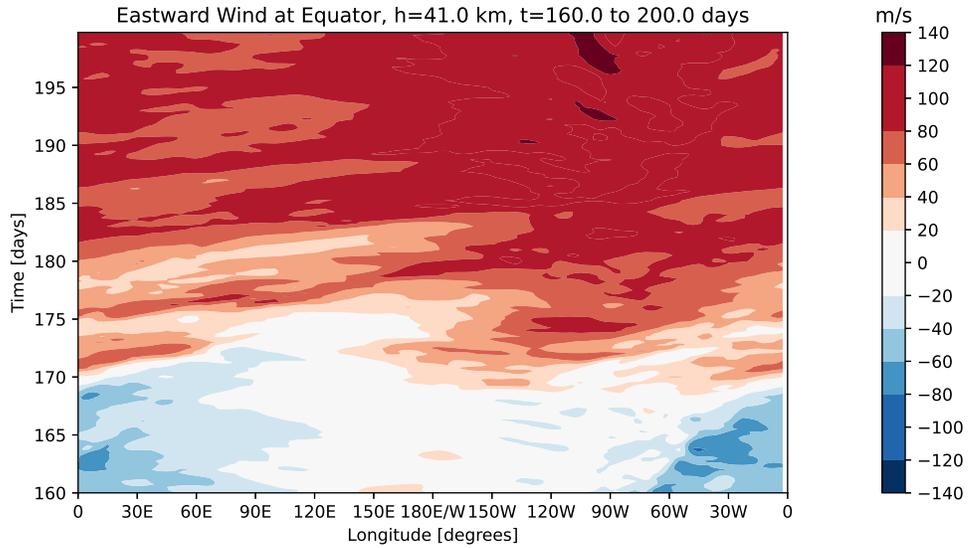}{0.9\textwidth}{(b) Transition from westward to eastward at 41 km}
\caption{Equatorial zonal wind speeds at 41~km over two 40-day periods, determined by six-hourly simulation data. The top panel shows the transition from eastward to westward winds and the bottom panel shows the transition from westward to eastward winds. Plots are centred on the antistellar point to highlight the tendency to resist westward flow in this region.}
\label{fig:transition}
\end{figure}

The zonal wind does not change direction simultaneously at every longitude. Figure \ref{fig:transition} shows the temporal variations of the 41~km zonal wind along the equator at two phase transitions, one from eastward to westward and the other from westward to eastward flow. In both transitions, longitudes close to the western terminator transition first, in addition to exhibiting higher wind speeds. Figure \ref{fig:transition}a shows that the westward winds begin to appear around 90$^\circ$W on days 125--130, while other longitudes transition around days 135--140. The converse is also true (Figure \ref{fig:transition}b). Wind speeds are consistently highest at 90$^\circ$--120$^\circ$W. Close inspection also shows that the antistellar point has a very weak westward phase, with wind speeds of only 5--10~ms$^{-1}$ in some places, compared to 100--120~ms$^{-1}$ during the eastward phase. Periods of westward flow at the antistellar point are comparatively brief. The weakness of the westward phase here implies lower westward acceleration, which may be due to the absorption of westward-propagating gravity waves before they travel around the planet to the nightside, a lack of critical levels available to absorb westward-propagating modes, an asymmetry in the underlying resolved gravity wave spectrum itself, or a combination of these mechanisms.

The amplitude of the LASO is largest at the western terminator and smallest at the antistellar point (Figure \ref{fig:transition}). This pattern cannot be explained directly by the presence of Rossby waves because the equatorial wind vector component contributed by the Rossby waves is larger at the antistellar point than at the western terminator. A possible explanation for the greater amplitude and early phase switching of the oscillation at the western terminator is the presence of a second gravity wave source in this region. The western terminator is a jet exit region where both stratospheric and tropospheric jets decrease rapidly in speed. The steep gradient in wind speeds is caused by the convergence of the two opposing wind vectors contributed by the Rossby wave cyclone-anticyclone pair. Jet exit regions are well-known to be sources of gravity waves \citep{plougonven_internal_2014}. To investigate further, we calculated the Lagrangian Rossby number (Eqn. \ref{eqn:rol}).

\begin{figure}

\gridline{\fig{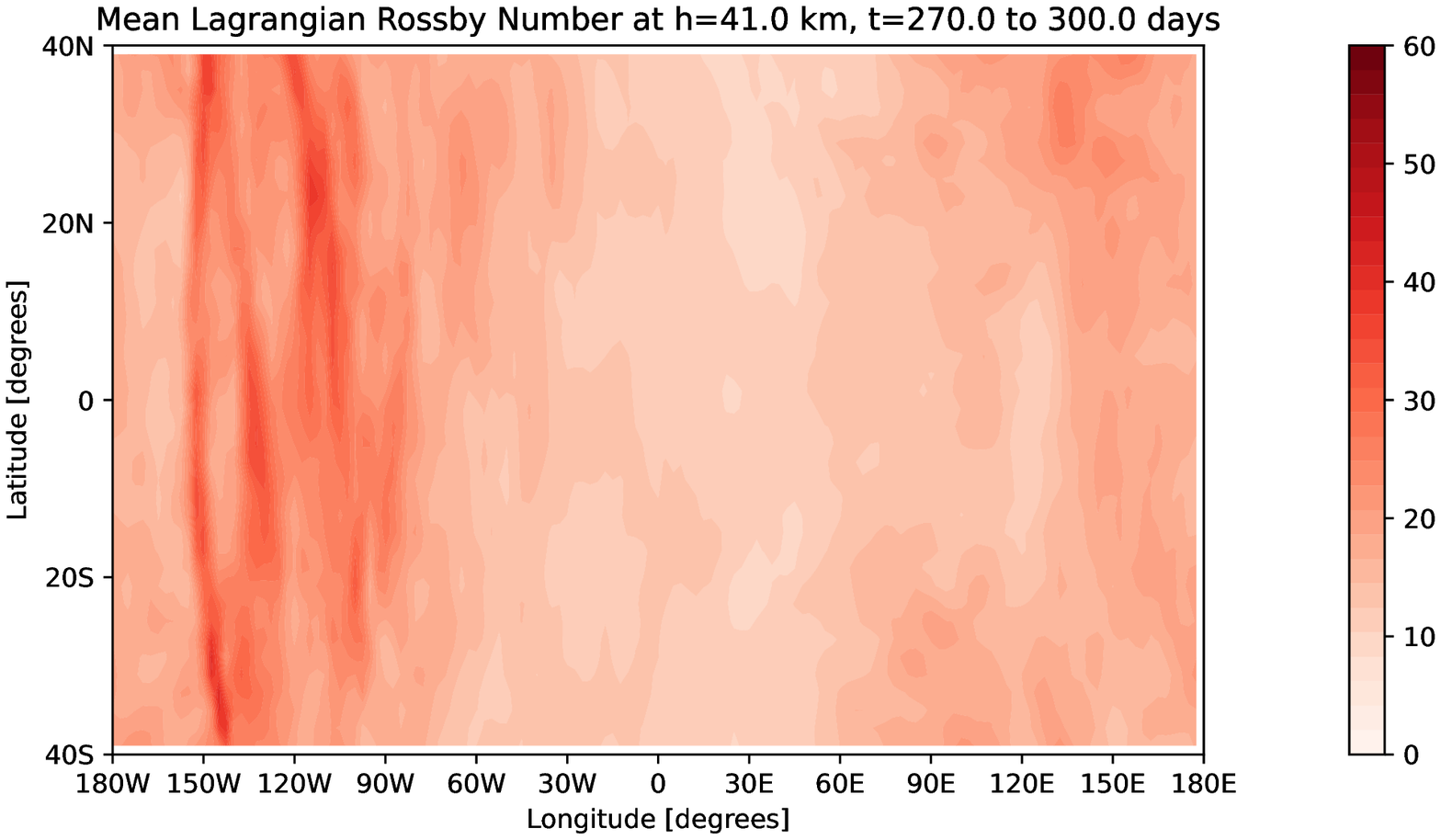}{0.5\textwidth}{(a)}\fig{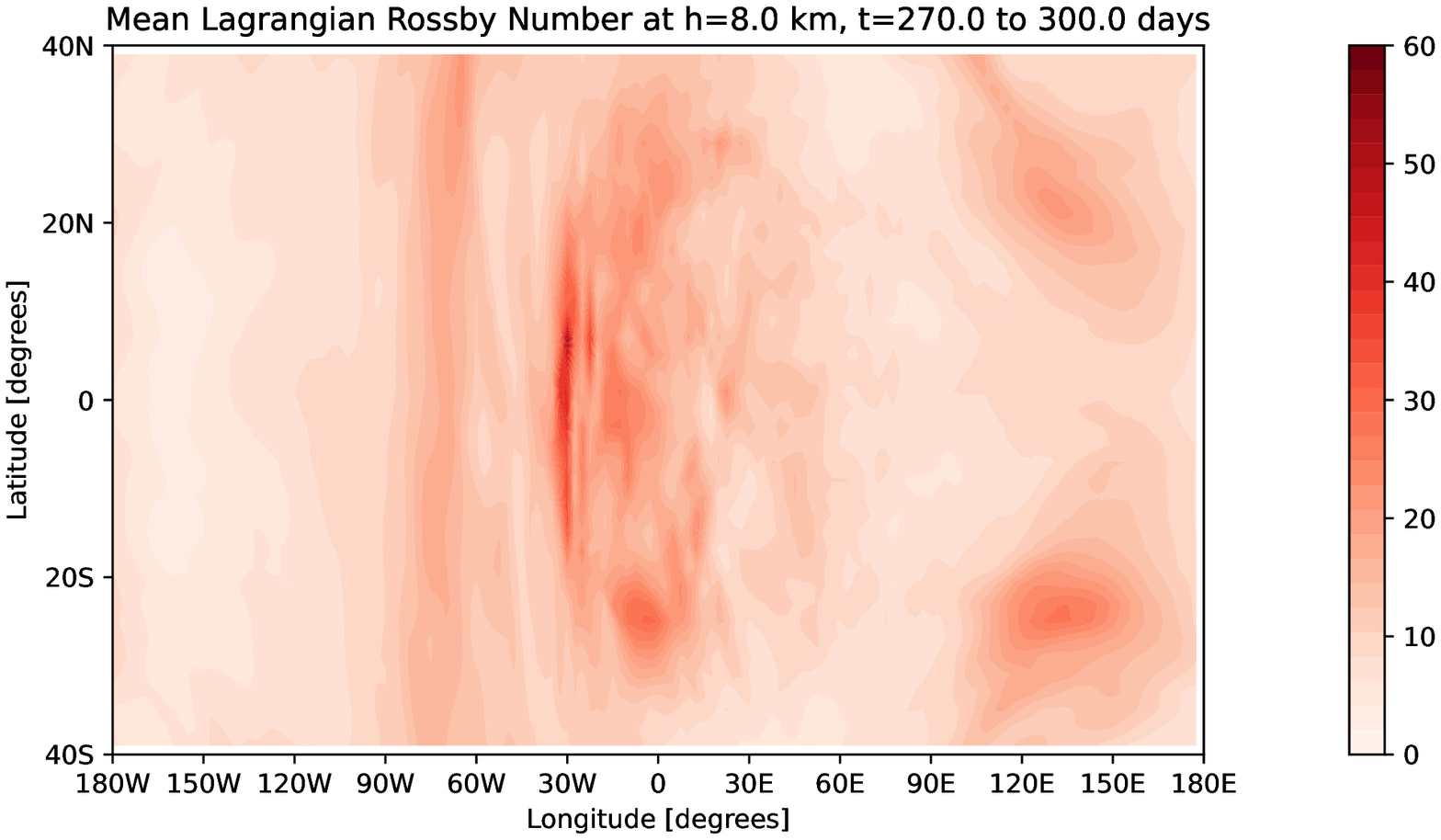}{0.5\textwidth}{(b)}}

\gridline{\fig{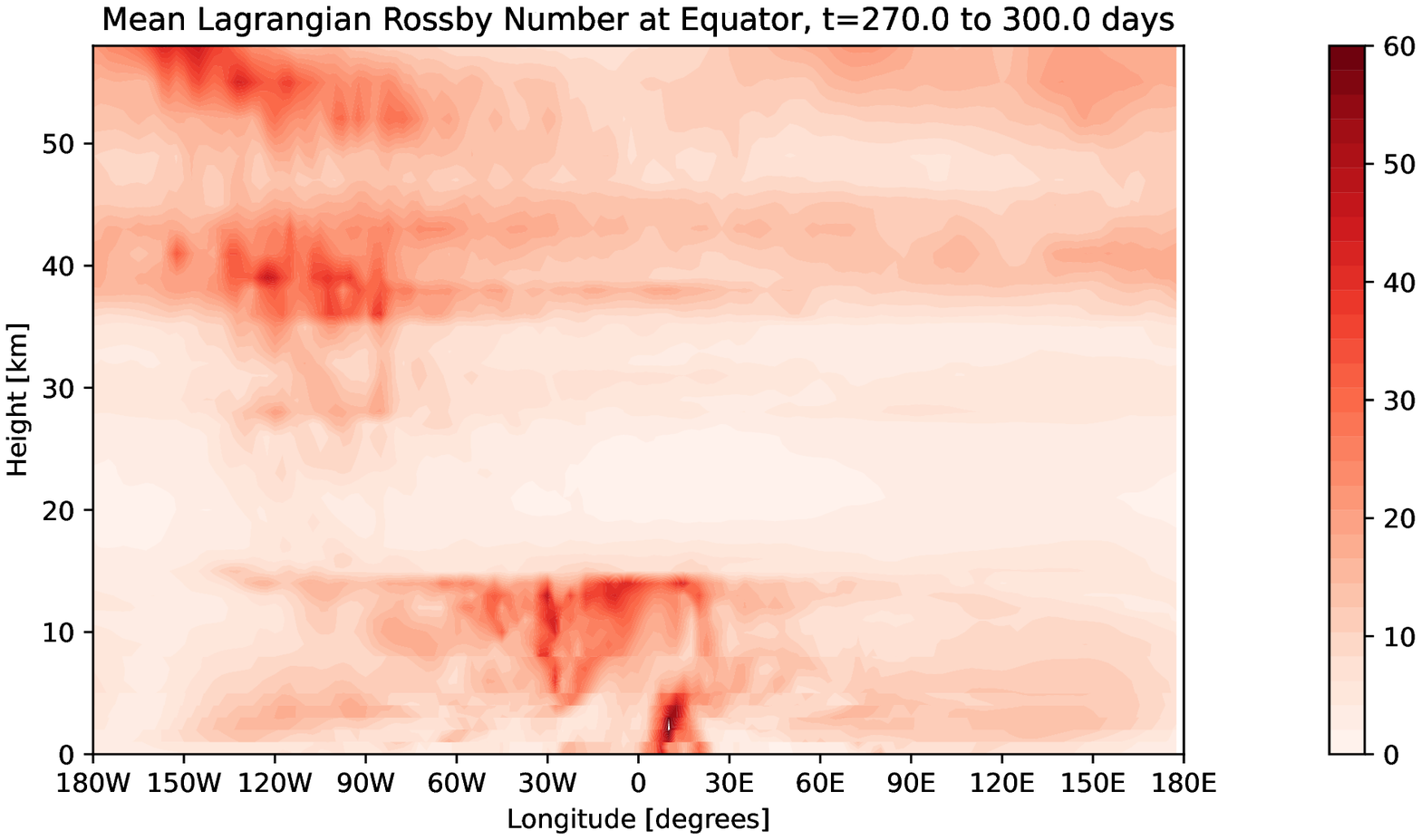}{0.5\textwidth}{(c)}}

\gridline{\fig{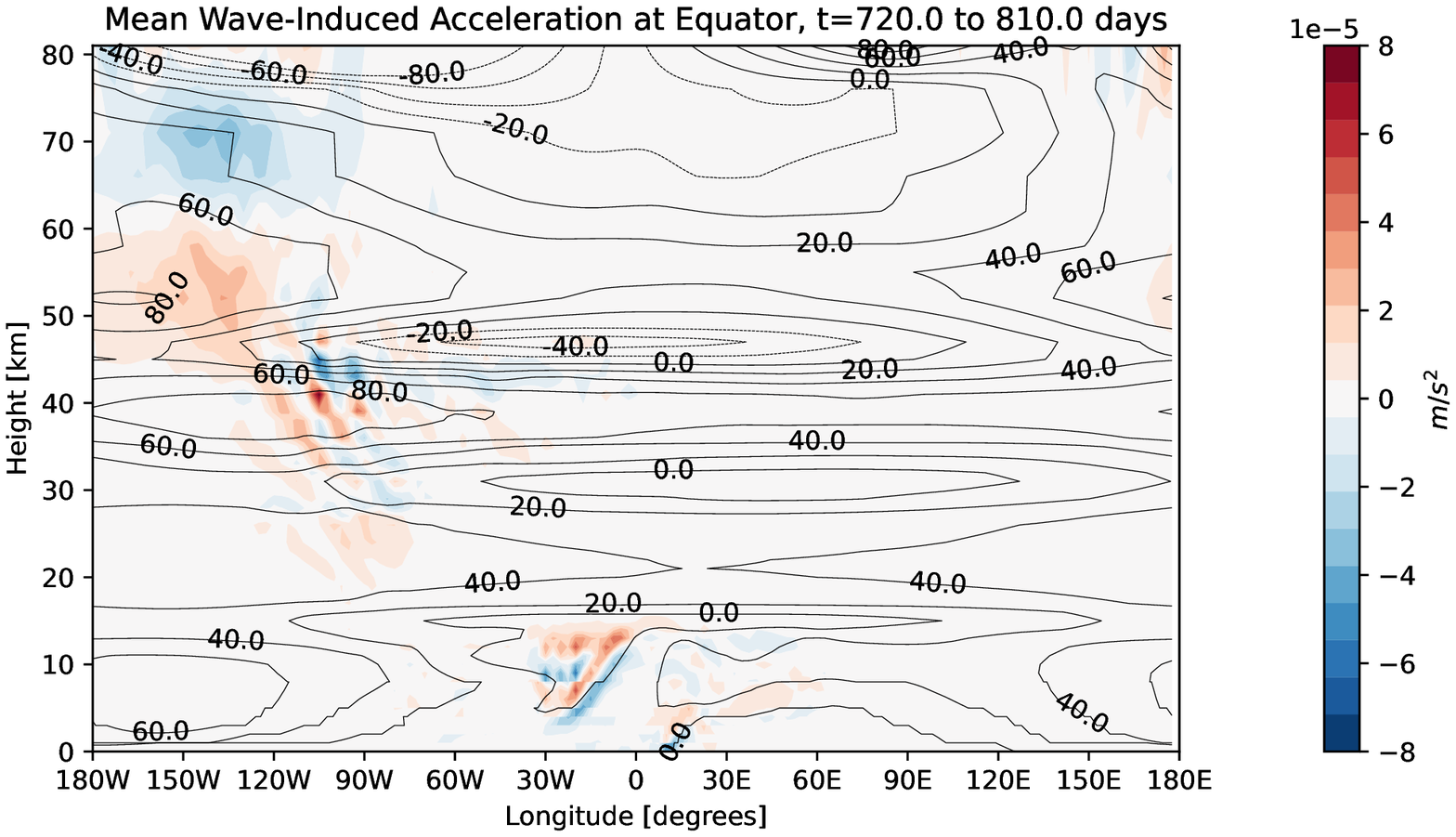}{1.0\textwidth}{(d)}}

\caption{The Lagrangian Rossby number, normalised by its minimum value in the dataset and averaged over the same 30-day period shown in Figure \ref{fig:deepconvection}, at a) 41~km and b) 8~km, and c) the corresponding height-longitude cross-section at the equator. We exclude the top five model levels and the high latitudes to avoid including model instability at the model top and poles. d) The height-longitude cross-section of gravity wave-induced acceleration over the same time period as Figure \ref{fig:waveacc}, overlaid with the mean zonal wind as contour lines.}
\label{fig:lrn}
\end{figure}

Figure \ref{fig:lrn}a shows that at $41$ km, where the LASO reaches its maximum amplitude, a strong gravity wave source is present at the western terminator. In the troposphere, unbalanced flows are found in the region of deep convection (Figure \ref{fig:lrn}b and Figure \ref{fig:deepconvection}). The equatorial altitude-longitude cross-section (Figure \ref{fig:lrn}c) confirms that areas of instability that can generate gravity waves occur preferentially westward of the substellar point in the troposphere and around the western terminator in the lower stratosphere. To link wave acceleration to the jet exit region, Figure \ref{fig:lrn}d shows the gravity wave-induced acceleration at the equator (as calculated in Figure \ref{fig:waveacc}), overlaid with the zonal wind. The strongest acceleration is found at the jet exit region for both the tropospheric and the lowest stratospheric jets between 20--45~km. The western terminator both generates gravity waves due to the slowing of the jet speeds in this region and also absorbs high-energy gravity waves because this is where the critical levels exhibit the highest wind speeds. The result is an unusually unstable region where zonal winds reverse earlier than elsewhere and may oscillate back and forth in shorter time periods, as seen in Figure \ref{fig:transition}a between 125--145 days and 120$^\circ$--90$^\circ$W .

We also ran the model using the experimental precipitation-dependent gravity wave scheme to investigate the impact of moving the gravity wave source on the characteristics of the LASO, as described in Section \ref{sec:methods}. In the precipitation-dependent scheme, parameterised gravity waves are launched preferentially from the convection zone around the substellar point on the dayside, creating an asymmetry in the parameterised gravity wave source that does not exist in the standard scheme. While the period, latitudinal extent, amplitude, resolved gravity waves, and water vapour oscillation did not differ between the precipitation-dependent and standard schemes, the longitudinal asymmetry in wind speeds and phase transitions \emph{decreased} and the LASO became more zonally symmetric. In the standard scheme, the wind velocity maximum at the western terminator and the minimum at approximately 30$^\circ$E differ by $\sim$100~ms$^{-1}$ at 41~km, while in the precipitation-dependent scheme they differ by only $\sim$20~ms$^{-1}$. The stronger gravity wave source at the substellar point appears to play a similar role as the source at the jet exit region in modifying the local flow at higher altitudes, counteracting the westward component contributed by the dayside cyclone. Although the overall mean circulation remains longitudinally asymmetrical, this experiment demonstrates the non-linearity in outcomes resulting from interactions between Rossby waves and gravity waves in the simulation.

\begin{figure}
    \plotone{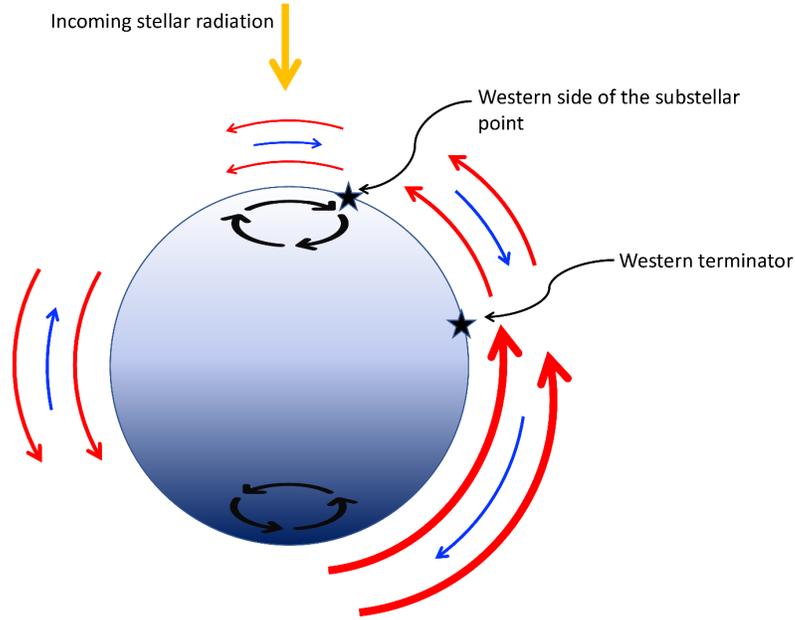}
    \caption{Schematic diagram of the location and interaction of Rossby waves, equatorial jets, and gravity wave sources. The perspective is looking downward from the north pole. The lighter side of the sphere is the dayside and the darker is the nightside. Red (blue) arrows indicate eastward (westward) air flow, with larger arrows representing faster wind speeds. Black spirals represent Rossby waves. The star symbol indicates gravity wave sources.}
    \label{fig:cartoon}
\end{figure}

In summary, as discussed above, the LASO is shaped by interactions between multiple components of the circulation typical of tidally locked planets. Figure \ref{fig:cartoon} shows a schematic illustration of the interplay between Rossby waves, the superrotating tropospheric jet, and the gravity wave sources in the substellar region and at the western terminators. The diagram represents the mean circulation generated with the standard, globally invariant gravity wave source.

\subsection{Atmospheric water vapour oscillations}
\label{subsec:vapour}

On Earth, the oscillation of the zonal wind direction is associated with variations in the amount of atmospheric water vapour of the order of $\pm$10\% \citep{randel_seasonal_1998, wang_tropical_2020}. An analogous oscillation on exoplanets is of interest because large fluctuations in atmospheric water vapour may be observable. In our simulations, specific humidity at 41~km varies about $\pm$50\% on average, with peak oscillations up to 200\% of the mean. This results in oscillations with minimum and maximum values that can differ by an order of magnitude. 

To test whether a fluctuation of this magnitude would be detectable, we generated synthetic planetary spectra as observed by the NIRSpec instrument aboard JWST for the above maximum and minimum water vapour concentrations. We found the fluctuations are too small to be observable for an Earth-sized planet like Proxima Centauri b, but as QBO-like phenomena have also been detected on Jupiter and Saturn in the solar system \citep{showman_atmospheric_2019, orton_thermal_1991, leovy_quasiquadrennial_1991}, a LASO could potentially occur on a larger planet with greater potential for observation. This could be particularly important for short-period, potentially tidally locked, super-Earth or mini-Neptune-type planets detected by missions such as the Transiting Exoplanet Survey Satellite (see \cite{bluhm_ultra-short-period_2021} for a recent example) 

\section{Discussion}\label{sec:discussion}

Using a global 3-D climate model we show that Earth-like tidally locked planets can exhibit a longitudinally asymmetric stratospheric wind oscillation (LASO), analogous to the quasi-biennial oscillation (QBO) on Earth. As QBO-like oscillations are also known to exist on Jupiter and Saturn \citep{showman_atmospheric_2019, orton_thermal_1991, leovy_quasiquadrennial_1991}, our finding suggests a similar conclusion for gas giants. The LASO reported in our study has implications for the climatology and observations of tidally locked exoplanets, as well as for our understanding of GCM results.

\subsection{Oscillations in meteorological and atmospheric composition}

It is well established that the QBO on Earth is responsible for oscillations in the abundance of many atmospheric species, notably ozone and methane \citep{bowman_global_1989, huang_ozone_2008, randel_seasonal_1998, li_what_2009}. In this initial study, we have focused on the properties and acceleration of the LASO. However, by including atmospheric chemical kinetics, future work could also explore LASO-induced variations in species important for observations. For example, Earth’s total ozone column fluctuates by $\pm$1~ppm ($\sim$ 4\%) on the 26--28 month timescale of the QBO. A variation of this magnitude is unlikely to be detectable with existing instrumentation, but an ozone column oscillation may be larger in amplitude on a tidally locked planet than on Earth, in line with our simulation results for water vapour. Studies of ozone photochemistry on tidally locked planets \citep{yates_ozone_2020} have also shown that the planet’s nightside accumulates ozone in its cold traps, resulting in a zonally asymmetrical ozone layer without analogy on Earth and high levels of ozone in one hemisphere. Future work could explore how this tidally locked ozone chemistry would interact with oscillations in dayside production of ozone through stratospheric photochemistry. In addition, LASO-related temperature anomalies in our simulation reach 40--50~K ($\sim$ 25\% deviation from the mean), while the QBO of temperature on Earth is limited to 1--2~K \citep{tegtmeier_zonal_2020}. Large variations in air temperature will also potentially affect measurements of thermal emission spectra from the atmosphere.

\subsection{Sensitivity to model parameterisations}

Gravity wave parametrisation schemes and differences in ability to generate a QBO-like phenomenon may account for some of the differences found in modelling studies of the same exoplanet. For example, predicted nighttime surface temperatures for Proxima Centauri b diverge by 50~K in different studies (e.g. \cite{turbet_habitability_2016, boutle_exploring_2017}), a discrepancy that can only be partly explained by differences in stellar irradiation used by these studies. \cite{boutle_exploring_2017} instead proposed that these temperature discrepancies could be explained by differences in convection parameterizations and model resolutions, which affect the water vapour profile and consequently the dayside-nightside heat transport. Gravity wave parameterisations likewise vary between models and \cite{watkins_gravity_2010} have shown that gravity waves can have a significant impact on the development of the background flow, temperature, and heat redistribution from dayside to nightside on tidally locked planets. Gravity waves can become trapped within jets and travel longitudinally from the gravity wave source on the dayside until the jet speed develops a critical level where the wave can be absorbed \citep{pitteway_reflection_1965, nappo_ducted_2012}. Differing gravity wave schemes may therefore result in different predictions for heat redistribution and dayside-nightside temperature contrast \citep{watkins_gravity_2010}. 

\subsection{The importance of the western terminator for atmospheric dynamics}

Our results identify the western terminator as a region of particular complexity. At the equatorial western terminator, the zonal wind components contributed by the standing Rossby waves on the dayside and nightside meet and oppose each other. We have also shown in Figure \ref{fig:lrn} that the western terminator is a jet exit region where the flow is unbalanced and internal gravity waves are generated and locally reabsorbed. These factors lead to significant turbulence and variability due to the interplay between these elements. As transit spectroscopy is expected to be a key tool in constraining the composition of exoplanet atmospheres, understanding the dynamics of the terminator regions of tidally locked planets is important to interpreting the results of observational studies. If the time scale of an oscillation is longer than the observing window, observers may view the planet in a low-temperature/high-temperature or low-abundance/high-abundance phase of an oscillation. As we noted above, a LASO on a hot Jupiter could be intense enough to be detectable. The high amplitude of the oscillation at the western terminator, a critical region for transit spectroscopy, suggests that the LASO should be studied on tidally locked hot Jupiters to determine its possible impact on observations.

\section{Conclusion}\label{sec:conclusion}

Using a global 3-D general circulation model, we reported the discovery of a phenomenon analogous to Earth's quasi-biennial oscillation (QBO) in simulations of the atmosphere of an Earth-like tidally locked planet, nominally Proxima Centauri b. As the phenomenon exhibits asymmetries related to the tidal locking, we refer to it as the longitudinally asymmetric stratospheric oscillation (LASO). The LASO begins above the tropospheric equatorial superrotating jet and extends vertically from roughly 35--55~km. It consists of a periodic reversal in the direction of the zonal winds within this region, accompanied by oscillations in temperature. We found the LASO in our simulation to have a latitudinal extent of $\pm$ 60$^\circ$ and a period of 5--6.5 Earth months. The shorter period compared to its Earth counterpart is explained to a first order by stronger wave driving from gravity wave generation in the deep convection zone at the substellar point. We also identified a secondary gravity wave source in the jet exit region at the western terminator. Our analysis found that wind speeds in the LASO reach a maximum at the western terminator and that reversals of direction at the western terminator precede those at other longitudes. We further found that the antistellar point experiences a very weak oscillation: zonal winds are preferentially eastward-flowing, with only brief and low-speed westward phases. 

Our results highlight the significance of atmospheric variability in global 3-D simulations of exoplanets and point towards a need for further studies to investigate this largely uncharted territory. In this initial study, we focused on the characteristics and mechanism of the LASO and did not explore the potential effect of oscillations in abundances of atmospheric species other than water vapour on exoplanet observations. Future work in this area could use a 3-D model with online atmospheric chemistry such as presented in \cite{yates_ozone_2020} to explore such effects. While we performed sensitivity tests by varying the horizontal and temporal resolution, we expect that the LASO's fundamental dependence on the planet's mean circulation will make it sensitive to factors that alter the planet's atmospheric dynamics, such as different atmospheric compositions, the presence of land, and the planet's rotation rate. On the Earth, the QBO is also known to interact with other dynamical phenomena such as the semi-annual oscillation, Madden-Julian oscillation \citep{martin_influence_2021}, and polar vortex. We leave such potential interactions between the LASO and similar sources of variability in the atmosphere for future work. 

As the body of observational data from exoplanets grows in the coming decades, our understanding of the parameter space for planetary climate states must grow with it. This parameter space includes time-dependent atmospheric phenomena such as the one we describe in this paper which, even if they are familiar features on Earth, will inevitably take new forms on tidally locked planets. Characterisation of the sources and properties of atmospheric variability on tidally locked exoplanets promises to refine our interpretations of observations and, in the case of terrestrial planets, deepen our understanding of the Earth in the context of the larger population of planets inside and outside the solar system.

\begin{acknowledgments}
\emph{Acknowledgments:}
Material produced using Met Office Software.

We gratefully acknowledge the funding and support provided by the Edinburgh Earth, Ecology, and Environmental Doctoral Training Partnership and the Natural Environmental Research Council [grant number NE/S007407/1]. We also kindly acknowledge our use of the Monsoon2 system, a collaborative facility supplied under the Joint Weather and Climate Research Programme, a strategic partnership between the Met Office and the Natural Environment Research Council. Our research was performed as part of the project `Using UKCA to investigate atmospheric composition on extra-solar planets (ExoChem).'

This work was partly funded by the Leverhulme Trust through a research project grant [RPG-2020-82], 
a Science and Technology Facilities Council Consolidated Grant [ST/R000395/1] and a UKRI Future Leaders Fellowship [grant number MR/T040866/1]. JM and IAB acknowledge the support of a Met Office Academic Partnership secondment. 
\end{acknowledgments}

%

\vspace{5mm}


\software{We used the Iris python package to manage and analyse model output data \citep{metoffice_iris_2010}. We used the windspharm python library to perform Helmholtz decompositions to characterise the mean circulation \citep{dawson_windspharm_2016} and NASA's Planetary Spectrum Generator to simulate transit spectra \citep{villanueva_planetary_2018}. 
          }




\bibliography{lasorefs}{}

\begin{thebibliography}{}
\expandafter\ifx\csname natexlab\endcsname\relax\def\natexlab#1{#1}\fi
\providecommand{\url}[1]{\href{#1}{#1}}
\providecommand{\dodoi}[1]{doi:~\href{http://doi.org/#1}{\nolinkurl{#1}}}
\providecommand{\doeprint}[1]{\href{http://ascl.net/#1}{\nolinkurl{http://ascl.net/#1}}}
\providecommand{\doarXiv}[1]{\href{https://arxiv.org/abs/#1}{\nolinkurl{https://arxiv.org/abs/#1}}}

\bibitem[{Anglada-Escudé {et~al.}(2016)Anglada-Escudé, Amado, Barnes,
  Berdiñas, Butler, Coleman, de~la Cueva, Dreizler, Endl, Giesers, Jeffers,
  Jenkins, Jones, Kiraga, Kürster, López-González, Marvin, Morales, Morin,
  Nelson, Ortiz, Ofir, Paardekooper, Reiners, Rodríguez, Rodrίguez-López,
  Sarmiento, Strachan, Tsapras, Tuomi, \&
  Zechmeister}]{anglada-escude_terrestrial_2016}
Anglada-Escudé, G., Amado, P.~J., Barnes, J., {et~al.} 2016, Nature, 536, 437,
  \dodoi{10.1038/nature19106}

\bibitem[{Arney {et~al.}(2017)Arney, Meadows, Domagal-Goldman, Deming,
  Robinson, Tovar, Wolf, \& Schwieterman}]{arney_pale_2017}
Arney, G.~N., Meadows, V.~S., Domagal-Goldman, S.~D., {et~al.} 2017, The
  Astrophysical Journal, 836, 49, \dodoi{10.3847/1538-4357/836/1/49}

\bibitem[{Baldwin {et~al.}(2001)Baldwin, Gray, Dunkerton, Hamilton, Haynes,
  Randel, Holton, Alexander, Hirota, Horinouchi, Jones, Kinnersley, Marquardt,
  Sato, \& Takahashi}]{baldwin_quasi-biennial_2001}
Baldwin, M.~P., Gray, L.~J., Dunkerton, T.~J., {et~al.} 2001, REVIEWS OF
  GEOPHYSICS, 52, \dodoi{https://doi.org/10.1029/1999RG000073}

\bibitem[{Barnes(2017)}]{barnes_tidal_2017}
Barnes, R. 2017, Celestial Mechanics and Dynamical Astronomy, 129, 509,
  \dodoi{10.1007/s10569-017-9783-7}

\bibitem[{Bluhm {et~al.}(2021)Bluhm, Palle, Molaverdikhani, Kemmer, Hatzes,
  Kossakowski, Stock, Caballero, Lillo-Box, Bejar, Soto, Amado, Brown, Cadieux,
  Cloutier, Collins, Collins, Cortes-Contreras, Doyon, Dreizler, Espinoza,
  Fukui, Gonzalez-Alvarez, Henning, Horne, Jeffers, Jenkins, Jensen, Kaminski,
  Kielkopf, Kusakabe, Kuerster, Lafreniere, Luque, Murgas, Montes, Morales,
  Narita, Passegger, Quirrenbach, Schoefer, Reffert, Reiners, Ribas, Ricker,
  Seager, Schweitzer, Schwarz, Tamura, Trifonov, Vanderspek, Winn, Zechmeister,
  \& Osorio}]{bluhm_ultra-short-period_2021}
Bluhm, P., Palle, E., Molaverdikhani, K., {et~al.} 2021, Astronomy \&
  Astrophysics, 650, A78, \dodoi{10.1051/0004-6361/202140688}

\bibitem[{Booker \& Bretherton(1967)}]{booker_critical_1967}
Booker, J.~R., \& Bretherton, F.~P. 1967, Journal of Fluid Mechanics, 27, 513,
  \dodoi{10.1017/S0022112067000515}

\bibitem[{Boutle {et~al.}(2020)Boutle, Joshi, Lambert, Mayne, Lyster, Manners,
  Ridgway, \& Kohary}]{boutle_mineral_2020}
Boutle, I.~A., Joshi, M., Lambert, F.~H., {et~al.} 2020, Nature Communications,
  11, 2731, \dodoi{10.1038/s41467-020-16543-8}

\bibitem[{Boutle {et~al.}(2017)Boutle, Mayne, Drummond, Manners, Goyal,
  Lambert, Acreman, \& Earnshaw}]{boutle_exploring_2017}
Boutle, I.~A., Mayne, N.~J., Drummond, B., {et~al.} 2017, Astronomy \&
  Astrophysics, 601, A120, \dodoi{10.1051/0004-6361/201630020}

\bibitem[{Bowman(1989)}]{bowman_global_1989}
Bowman, K.~P. 1989, Journal of the Atmospheric Sciences, 46, 3328,
  \dodoi{10.1175/1520-0469(1989)046<3328:GPOTQB>2.0.CO;2}

\bibitem[{Braesicke(2015)}]{braesicke_middle_2015}
Braesicke, P. 2015, in Encyclopedia of {Atmospheric} {Sciences} ({Second}
  {Edition}), ed. G.~R. North, J.~Pyle, \& F.~Zhang (Oxford: Academic Press),
  50--56, \dodoi{10.1016/B978-0-12-382225-3.00227-9}

\bibitem[{Brogi {et~al.}(2016)Brogi, Kok, Albrecht, Snellen, Birkby, \&
  Schwarz}]{brogi_rotation_2016}
Brogi, M., Kok, R. J.~d., Albrecht, S., {et~al.} 2016, The Astrophysical
  Journal, 817, 106, \dodoi{10.3847/0004-637X/817/2/106}

\bibitem[{Bushell(2021)}]{bushell_umdp_2021}
Bushell, A. 2021, {UMDP} 034: {Non}-{Orographic} ({Spectral}) {Gravity} {Wave}
  {Parametrization}, Tech. Rep.~34, Met Office, Exeter, United Kingdom.
\newblock \url{https://code.metoffice.gov.uk/doc/um/latest/papers/umdp_034.pdf}

\bibitem[{Bushell {et~al.}(2015)Bushell, Butchart, Derbyshire, Jackson, Shutts,
  Vosper, \& Webster}]{bushell_parameterized_2015}
Bushell, A.~C., Butchart, N., Derbyshire, S.~H., {et~al.} 2015, Journal of the
  Atmospheric Sciences, 72, 4349, \dodoi{10.1175/JAS-D-15-0022.1}

\bibitem[{Dawson(2016)}]{dawson_windspharm_2016}
Dawson, A. 2016, Journal of Open Research Software, 4, e31,
  \dodoi{10.5334/jors.129}

\bibitem[{Debras {et~al.}(2019)Debras, Mayne, Baraffe, Goffrey, \&
  Thuburn}]{debras_eigenvectors_2019}
Debras, F., Mayne, N., Baraffe, I., Goffrey, T., \& Thuburn, J. 2019, Astronomy
  \& Astrophysics, 631, A36, \dodoi{10.1051/0004-6361/201935582}

\bibitem[{Debras {et~al.}(2020)Debras, Mayne, Baraffe, Jaupart, Mourier, Laibe,
  Goffrey, \& Thuburn}]{debras_acceleration_2020}
Debras, F., Mayne, N., Baraffe, I., {et~al.} 2020, Astronomy \& Astrophysics,
  633, A2, \dodoi{10.1051/0004-6361/201936110}

\bibitem[{Demory {et~al.}(2016)Demory, Gillon, de~Wit, Madhusudhan, Bolmont,
  Heng, Kataria, Lewis, Hu, Krick, Stamenkovic, Benneke, Kane, \&
  Queloz}]{demory_map_2016}
Demory, B.-O., Gillon, M., de~Wit, J., {et~al.} 2016, Nature, 532, 207,
  \dodoi{10.1038/nature17169}

\bibitem[{Dunkerton(1997)}]{dunkerton_role_1997}
Dunkerton, T.~J. 1997, Journal of Geophysical Research: Atmospheres, 102,
  26053, \dodoi{https://doi.org/10.1029/96JD02999}

\bibitem[{Dunkerton {et~al.}(2015)Dunkerton, Anstey, \&
  Gray}]{dunkerton_middle_2015}
Dunkerton, T.~J., Anstey, J.~A., \& Gray, L.~J. 2015, in Encyclopedia of
  {Atmospheric} {Sciences} ({Second} {Edition}), ed. G.~R. North, J.~Pyle, \&
  F.~Zhang (Oxford: Academic Press), 18--25,
  \dodoi{10.1016/B978-0-12-382225-3.00232-2}

\bibitem[{Dunkerton \& Delisi(1985)}]{dunkerton_climatology_1985}
Dunkerton, T.~J., \& Delisi, D.~P. 1985, Journal of the Atmospheric Sciences,
  42, 376, \dodoi{10.1175/1520-0469(1985)042<0376:COTELS>2.0.CO;2}

\bibitem[{Fauchez {et~al.}(2021)Fauchez, Villanueva, Sergeev, Turbet, Boutle,
  Tsigaridis, Way, Wolf, Domagal-Goldman, Forget, Haqq-Misra, Kopparapu,
  Manners, \& Mayne}]{fauchez_trappist-1_2021}
Fauchez, T.~J., Villanueva, G.~L., Sergeev, D.~E., {et~al.} 2021,
  arXiv:2109.11460 [astro-ph, physics:physics].
\newblock \url{http://arxiv.org/abs/2109.11460}

\bibitem[{Frierson {et~al.}(2006)Frierson, Held, \&
  Zurita-Gotor}]{frierson_gray-radiation_2006}
Frierson, D. M.~W., Held, I.~M., \& Zurita-Gotor, P. 2006, Journal of
  Atmospheric Sciences, 63, 2548, \dodoi{10.1175/JAS3753.1}

\bibitem[{Garfinkel {et~al.}(2021)Garfinkel, Gerber, Shamir, Rao, Jucker,
  White, \& Paldor}]{garfinkel_qbo_2021}
Garfinkel, C.~I., Gerber, E.~P., Shamir, O., {et~al.} 2021, A {QBO} cookbook:
  {Sensitivity} of the {Quasi}-{Biennial} {Oscillation} to resolution, resolved
  waves, and parameterized gravity waves, \dodoi{10.1002/essoar.10506774.1}

\bibitem[{Hammond \& Lewis(2021)}]{hammond_rotational_2021}
Hammond, M., \& Lewis, N.~T. 2021, Proceedings of the National Academy of
  Sciences, 118, \dodoi{10.1073/pnas.2022705118}

\bibitem[{Haynes(1998)}]{haynes_latitudinal_1998}
Haynes, P.~H. 1998, Quarterly Journal of the Royal Meteorological Society, 124,
  2645, \dodoi{https://doi.org/10.1002/qj.49712455206}

\bibitem[{Heath {et~al.}(1999)Heath, Doyle, Joshi, \&
  Haberle}]{heath_habitability_1999}
Heath, M.~J., Doyle, L.~R., Joshi, M.~M., \& Haberle, R.~M. 1999, Origins of
  life and evolution of the biosphere, 29, 405, \dodoi{10.1023/A:1006596718708}

\bibitem[{Huang {et~al.}(2008)Huang, Mayr, Reber, Russell, Mlynczak, \&
  Mengel}]{huang_ozone_2008}
Huang, F.~T., Mayr, H.~G., Reber, C.~A., {et~al.} 2008, Journal of Geophysical
  Research: Space Physics, 113, \dodoi{10.1029/2007JA012634}

\bibitem[{Joshi {et~al.}(1997)Joshi, Haberle, \&
  Reynolds}]{joshi_simulations_1997}
Joshi, M.~M., Haberle, R.~M., \& Reynolds, R.~T. 1997, Icarus, 129, 450,
  \dodoi{10.1006/icar.1997.5793}

\bibitem[{Koll \& Abbot(2016)}]{koll_temperature_2016}
Koll, D. D.~B., \& Abbot, D.~S. 2016, The Astrophysical Journal, 825, 99,
  \dodoi{10.3847/0004-637X/825/2/99}

\bibitem[{Komacek \& Abbot(2019)}]{komacek_atmospheric_2019}
Komacek, T.~D., \& Abbot, D.~S. 2019, The Astrophysical Journal, 871, 245,
  \dodoi{10.3847/1538-4357/aafb33}

\bibitem[{Kopparapu(2013)}]{kopparapu_revised_2013}
Kopparapu, R.~K. 2013, The Astrophysical Journal, 767, L8,
  \dodoi{10.1088/2041-8205/767/1/L8}

\bibitem[{Kopparapu {et~al.}(2016)Kopparapu, Wolf, Haqq-Misra, Yang, Kasting,
  Meadows, Terrien, \& Mahadevan}]{kopparapu_inner_2016}
Kopparapu, R.~k., Wolf, E.~T., Haqq-Misra, J., {et~al.} 2016, The Astrophysical
  Journal, 819, 84, \dodoi{10.3847/0004-637X/819/1/84}

\bibitem[{Labonté \& Merlis(2020)}]{labonte_sensitivity_2020}
Labonté, M.-P., \& Merlis, T.~M. 2020, The Astrophysical Journal, 896, 31,
  \dodoi{10.3847/1538-4357/ab9102}

\bibitem[{Lane(2015)}]{lane_gravity_2015}
Lane, T.~P. 2015, in Encyclopedia of {Atmospheric} {Sciences} ({Second}
  {Edition}), ed. G.~R. North, J.~Pyle, \& F.~Zhang (Oxford: Academic Press),
  171--179, \dodoi{10.1016/B978-0-12-382225-3.00489-8}

\bibitem[{Leovy {et~al.}(1991)Leovy, Friedson, \&
  Orton}]{leovy_quasiquadrennial_1991}
Leovy, C.~B., Friedson, A.~J., \& Orton, G.~S. 1991, Nature, 354, 380,
  \dodoi{10.1038/354380a0}

\bibitem[{Li {et~al.}(2009)Li, Palmer, Pumphrey, Bernath, \&
  Mahieu}]{li_what_2009}
Li, Q., Palmer, P.~I., Pumphrey, H.~C., Bernath, P., \& Mahieu, E. 2009,
  Atmospheric Chemistry and Physics, 9, 8531, \dodoi{10.5194/acp-9-8531-2009}

\bibitem[{Lin(2007)}]{lin_mesoscale_2007}
Lin, Y.-L. 2007, Mesoscale {Dynamics} (Cambridge: Cambridge University Press),
  \dodoi{10.1017/CBO9780511619649}

\bibitem[{Lindzen \& Holton(1968)}]{lindzen_theory_1968}
Lindzen, R., \& Holton, J. 1968, J. Atmos. Sci., 25, 1095,
  \dodoi{10.1175/1520-0469(1968)025<1095:ATOTQB>2.0.CO;2}

\bibitem[{Lines {et~al.}(2018)Lines, Manners, Mayne, Goyal, Carter, Boutle,
  Lee, Helling, Drummond, Acreman, \& Sing}]{lines_exonephology_2018}
Lines, S., Manners, J., Mayne, N.~J., {et~al.} 2018, Monthly Notices of the
  Royal Astronomical Society, 481, 194, \dodoi{10.1093/mnras/sty2275}

\bibitem[{Louden \& Wheatley(2015)}]{louden_spatially_2015}
Louden, T., \& Wheatley, P.~J. 2015, The Astrophysical Journal, 814, L24,
  \dodoi{10.1088/2041-8205/814/2/L24}

\bibitem[{Martin {et~al.}(2021)Martin, Son, Butler, Hendon, Kim, Sobel, Yoden,
  \& Zhang}]{martin_influence_2021}
Martin, Z., Son, S.-W., Butler, A., {et~al.} 2021, Nature Reviews Earth \&
  Environment, 1, \dodoi{10.1038/s43017-021-00173-9}

\bibitem[{May {et~al.}(2021)May, Taylor, Komacek, Line, \&
  Parmentier}]{may_water_2021}
May, E.~M., Taylor, J., Komacek, T.~D., Line, M.~R., \& Parmentier, V. 2021,
  arXiv:2103.09313 [astro-ph].
\newblock \url{http://arxiv.org/abs/2103.09313}

\bibitem[{Mayne {et~al.}(2014)Mayne, Baraffe, Acreman, Smith, Wood, Amundsen,
  Thuburn, \& Jackson}]{mayne_using_2014}
Mayne, N.~J., Baraffe, I., Acreman, D.~M., {et~al.} 2014, Geoscientific Model
  Development, 7, 3059, \dodoi{10.5194/gmd-7-3059-2014}

\bibitem[{Merlis \& Schneider(2010)}]{merlis_atmospheric_2010}
Merlis, T.~M., \& Schneider, T. 2010, Journal of Advances in Modeling Earth
  Systems, 2, \dodoi{10.3894/JAMES.2010.2.13}

\bibitem[{MetOffice(2010)}]{metoffice_iris_2010}
MetOffice. 2010, Iris: {A} {Python} package for analysing and visualising
  meteorological and oceanographic data sets.
\newblock \url{https://scitools-iris.readthedocs.io/en/stable/}

\bibitem[{Mollière {et~al.}(2017)Mollière, Boekel, Bouwman, Henning, Lagage,
  \& Min}]{molliere_observing_2017}
Mollière, P., Boekel, R.~v., Bouwman, J., {et~al.} 2017, Astronomy \&
  Astrophysics, 600, A10, \dodoi{10.1051/0004-6361/201629800}

\bibitem[{Morley {et~al.}(2017)Morley, Kreidberg, Rustamkulov, Robinson, \&
  Fortney}]{morley_observing_2017}
Morley, C.~V., Kreidberg, L., Rustamkulov, Z., Robinson, T., \& Fortney, J.~J.
  2017, The Astrophysical Journal, 850, 121, \dodoi{10.3847/1538-4357/aa927b}

\bibitem[{Nappo(2012)}]{nappo_ducted_2012}
Nappo, C.~J. 2012, in An {Introduction} to {Atmospheric} {Gravity} {Waves},
  Vol. 102, International {Geophysics}, ed. C.~J. Nappo (Academic Press),
  87--116, \dodoi{10.1016/B978-0-12-385223-6.00004-5}

\bibitem[{Orton {et~al.}(1991)Orton, Friedson, Baines, Martin, West, Caldwell,
  Hammel, Bergstralh, Malcom, Golisch, Griep, Kaminski, Tokunaga, Baron, \&
  Shure}]{orton_thermal_1991}
Orton, G.~S., Friedson, A.~J., Baines, K.~H., {et~al.} 1991, Science, 252, 537,
  \dodoi{10.1126/science.252.5005.537}

\bibitem[{Pierrehumbert \& Hammond(2019)}]{pierrehumbert_atmospheric_2019}
Pierrehumbert, R.~T., \& Hammond, M. 2019, Annual Review of Fluid Mechanics,
  51, 275, \dodoi{10.1146/annurev-fluid-010518-040516}

\bibitem[{Pitteway \& Hines(1965)}]{pitteway_reflection_1965}
Pitteway, M. L.~V., \& Hines, C.~O. 1965, Canadian Journal of Physics, 43,
  2222, \dodoi{10.1139/p65-217}

\bibitem[{Plougonven \& Zhang(2014)}]{plougonven_internal_2014}
Plougonven, R., \& Zhang, F. 2014, Reviews of Geophysics, 52, 33,
  \dodoi{10.1002/2012RG000419}

\bibitem[{Plumb(1977)}]{plumb_interaction_1977}
Plumb, R.~A. 1977, Journal of the Atmospheric Sciences, 34, 1847,
  \dodoi{10.1175/1520-0469(1977)034<1847:TIOTIW>2.0.CO;2}

\bibitem[{Plumb \& McEwan(1978)}]{plumb_instability_1978}
Plumb, R.~A., \& McEwan, A.~D. 1978, Journal of the Atmospheric Sciences, 35,
  1827, \dodoi{10.1175/1520-0469(1978)035<1827:TIOAFS>2.0.CO;2}

\bibitem[{Rajpurohit {et~al.}(2013)Rajpurohit, Reylé, Allard, Homeier,
  Schultheis, Bessell, \& Robin}]{rajpurohit_effective_2013}
Rajpurohit, A.~S., Reylé, C., Allard, F., {et~al.} 2013, Astronomy \&
  Astrophysics, 556, A15, \dodoi{10.1051/0004-6361/201321346}

\bibitem[{Randel {et~al.}(1998)Randel, Wu, Russell, Roche, \&
  Waters}]{randel_seasonal_1998}
Randel, W.~J., Wu, F., Russell, J.~M., Roche, A., \& Waters, J.~W. 1998,
  Journal of the Atmospheric Sciences, 55, 163,
  \dodoi{10.1175/1520-0469(1998)055<0163:SCAQVI>2.0.CO;2}

\bibitem[{Scaife {et~al.}(2000)Scaife, Butchart, Warner, Stainforth, Norton, \&
  Austin}]{scaife_realistic_2000}
Scaife, A.~A., Butchart, N., Warner, C.~D., {et~al.} 2000, Geophysical Research
  Letters, 27, 3481, \dodoi{https://doi.org/10.1029/2000GL011625}

\bibitem[{Scaife {et~al.}(2002)Scaife, Butchart, Warner, \&
  Swinbank}]{scaife_impact_2002}
Scaife, A.~A., Butchart, N., Warner, C.~D., \& Swinbank, R. 2002, Journal of
  the Atmospheric Sciences, 59, 1473,
  \dodoi{10.1175/1520-0469(2002)059<1473:IOASGW>2.0.CO;2}

\bibitem[{Scalo {et~al.}(2007)Scalo, Kaltenegger, Segura, Fridlund, Ribas,
  Kulikov, Grenfell, Rauer, Odert, Leitzinger, Selsis, Khodachenko, Eiroa,
  Kasting, \& Lammer}]{scalo_m_2007}
Scalo, J., Kaltenegger, L., Segura, A.~G., {et~al.} 2007, Astrobiology, 7, 85,
  \dodoi{10.1089/ast.2006.0000}

\bibitem[{Schenzinger {et~al.}(2017)Schenzinger, Osprey, Gray, \&
  Butchart}]{schenzinger_defining_2017}
Schenzinger, V., Osprey, S., Gray, L., \& Butchart, N. 2017, Geoscientific
  Model Development, 10, 2157, \dodoi{10.5194/gmd-10-2157-2017}

\bibitem[{Schwieterman {et~al.}(2018)Schwieterman, Kiang, Parenteau, Harman,
  DasSarma, Fisher, Arney, Hartnett, Reinhard, Olson, Meadows, Cockell, Walker,
  Grenfell, Hegde, Rugheimer, Hu, \& Lyons}]{schwieterman_exoplanet_2018}
Schwieterman, E.~W., Kiang, N.~Y., Parenteau, M.~N., {et~al.} 2018,
  Astrobiology, 18, 663, \dodoi{10.1089/ast.2017.1729}

\bibitem[{Sergeev {et~al.}(2020)Sergeev, Lambert, Mayne, Boutle, Manners, \&
  Kohary}]{sergeev_atmospheric_2020}
Sergeev, D.~E., Lambert, F.~H., Mayne, N.~J., {et~al.} 2020, Astrophysical
  Journal, 894, 84, \dodoi{10.3847/1538-4357/ab8882}

\bibitem[{Sergeev {et~al.}(2021)Sergeev, Fauchez, Turbet, Boutle, Tsigaridis,
  Way, Wolf, Domagal-Goldman, Forget, Haqq-Misra, Kopparapu, Lambert, Manners,
  \& Mayne}]{sergeev_trappist-1_2021}
Sergeev, D.~E., Fauchez, T.~J., Turbet, M., {et~al.} 2021, arXiv:2109.11459
  [astro-ph, physics:physics].
\newblock \url{http://arxiv.org/abs/2109.11459}

\bibitem[{Showman \& Polvani(2010)}]{showman_matsuno-gill_2010}
Showman, A.~P., \& Polvani, L.~M. 2010, Geophysical Research Letters, 37,
  \dodoi{https://doi.org/10.1029/2010GL044343}

\bibitem[{Showman \& Polvani(2011)}]{showman_equatorial_2011}
---. 2011, The Astrophysical Journal, 738, 71,
  \dodoi{10.1088/0004-637X/738/1/71}

\bibitem[{Showman {et~al.}(2019)Showman, Tan, \&
  Zhang}]{showman_atmospheric_2019}
Showman, A.~P., Tan, X., \& Zhang, X. 2019, The Astrophysical Journal, 883, 4,
  \dodoi{10.3847/1538-4357/ab384a}

\bibitem[{Showman {et~al.}(2013)Showman, Wordsworth, Merlis, \&
  Kaspi}]{showman_atmospheric_2013}
Showman, A.~P., Wordsworth, R.~D., Merlis, T.~M., \& Kaspi, Y. 2013, in
  Comparative {Climatology} of {Terrestrial} {Planets} (University of Arizona
  Press).
\newblock \url{https://arxiv.org/pdf/1306.2418.pdf}

\bibitem[{Sugimoto {et~al.}(2021)Sugimoto, Fujisawa, Kashimura, Noguchi,
  Kuroda, Takagi, \& Hayashi}]{sugimoto_generation_2021}
Sugimoto, N., Fujisawa, Y., Kashimura, H., {et~al.} 2021, Nature
  Communications, 12, 3682, \dodoi{10.1038/s41467-021-24002-1}

\bibitem[{Suissa {et~al.}(2020)Suissa, Mandell, Wolf, Villanueva, Fauchez, \&
  Kopparapu}]{suissa_dim_2020}
Suissa, G., Mandell, A.~M., Wolf, E.~T., {et~al.} 2020, The Astrophysical
  Journal, 891, 58, \dodoi{10.3847/1538-4357/ab72f9}

\bibitem[{Tegtmeier {et~al.}(2020)Tegtmeier, Anstey, Davis, Ivanciu, Jia,
  McPhee, \& Kedzierski}]{tegtmeier_zonal_2020}
Tegtmeier, S., Anstey, J., Davis, S., {et~al.} 2020, Geophysical Research
  Letters, 47, e2020GL089533, \dodoi{10.1029/2020GL089533}

\bibitem[{Tinetti {et~al.}(2018)Tinetti, Drossart, Eccleston, Hartogh, Heske,
  Leconte, Micela, Ollivier, Pilbratt, Puig, Turrini, Vandenbussche,
  Wolkenberg, Beaulieu, Buchave, Ferus, Griffin, Guedel, Justtanont, Lagage,
  Machado, Malaguti, Min, Norgaard-Nielsen, Rataj, Ray, Ribas, Swain, Szabo,
  Werner, Barstow, Burleigh, Cho, du~Foresto, Coustenis, Decin, Encrenaz,
  Galand, Gillon, Helled, Carlos~Morales, Munoz, Moneti, Pagano, Pascale,
  Piccioni, Pinfield, Sarkar, Selsis, Tennyson, Triaud, Venot, Waldmann,
  Waltham, Wright, Amiaux, Augueres, Berthe, Bezawada, Bishop, Bowles, Coffey,
  Colome, Crook, Crouzet, Da~Peppo, Sanz, Focardi, Frericks, Hunt, Kohley,
  Middleton, Morgante, Ottensamer, Pace, Pearson, Stamper, Symonds, Rengel,
  Renotte, Ade, Affer, Alard, Allard, Altieri, Andre, Arena, Argyriou, Aylward,
  Baccani, Bakos, Banaszkiewicz, Barlow, Batista, Bellucci, Benatti, Bernardi,
  Bezard, Blecka, Bolmont, Bonfond, Bonito, Bonomo, Brucato, Brun, Bryson,
  Bujwan, Casewell, Charnay, Pestellini, Chen, Ciaravella, Claudi, Cledassou,
  Damasso, Damiano, Danielski, Deroo, Di~Giorgio, Dominik, Doublier, Doyle,
  Doyon, Drummond, Duong, Eales, Edwards, Farina, Flaccomio, Fletcher, Forget,
  Fossey, Fraenz, Fujii, Garcia-Piquer, Gear, Geoffray, Gerard, Gesa, Gomez,
  Graczyk, Griffith, Grodent, Guarcello, Gustin, Hamano, Hargrave, Hello, Heng,
  Herrero, Hornstrup, Hubert, Ida, Ikoma, Iro, Irwin, Jarchow, Jaubert, Jones,
  Julien, Kameda, Kerschbaum, Kervella, Koskinen, Krijger, Krupp, Lafarga,
  Landini, Lellouch, Leto, Luntzer, Rank-Luftinger, Maggio, Maldonado,
  Maillard, Mall, Marquette, Mathis, Maxted, Matsuo, Medvedev, Miguel, Minier,
  Morello, Mura, Narita, Nascimbeni, Nguyen~Tong, Noce, Oliva, Palle, Palmer,
  Pancrazzi, Papageorgiou, Parmentier, Perger, Petralia, Pezzuto,
  Pierrehumbert, Pillitteri, Piotto, Pisano, Prisinzano, Radioti, Reess, Rezac,
  Rocchetto, Rosich, Sanna, Santerne, Savini, Scandariato, Sicardy, Sierra,
  Sindoni, Skup, Snellen, Sobiecki, Soret, Sozzetti, Stiepen, Strugarek,
  Taylor, Taylor, Terenzi, Tessenyi, Tsiaras, Tucker, Valencia, Vasisht, Vazan,
  Vilardell, Vinatier, Viti, Waters, Wawer, Wawrzaszek, Whitworth, Yung,
  Yurchenko, Zapatero~Osorio, Zellem, Zingales, \&
  Zwart}]{tinetti_chemical_2018}
Tinetti, G., Drossart, P., Eccleston, P., {et~al.} 2018, Experimental
  Astronomy, 46, 135, \dodoi{10.1007/s10686-018-9598-x}

\bibitem[{Tsai {et~al.}(2014)Tsai, Dobbs-Dixon, \&
  Gu}]{tsai_three-dimensional_2014}
Tsai, S.-M., Dobbs-Dixon, I., \& Gu, P.-G. 2014, The Astrophysical Journal,
  793, 141, \dodoi{10.1088/0004-637X/793/2/141}

\bibitem[{Turbet {et~al.}(2016)Turbet, Leconte, Selsis, Bolmont, Forget, Ribas,
  Raymond, \& Anglada-Escude}]{turbet_habitability_2016}
Turbet, M., Leconte, J., Selsis, F., {et~al.} 2016, Astronomy \& Astrophysics,
  596, A112, \dodoi{10.1051/0004-6361/201629577}

\bibitem[{Turbet {et~al.}(2021)Turbet, Fauchez, Sergeev, Boutle, Tsigaridis,
  Way, Wolf, Domagal-Goldman, Forget, Haqq-Misra, Kopparapu, Lambert, Manners,
  Mayne, \& Sohl}]{turbet_trappist-1_2021}
Turbet, M., Fauchez, T.~J., Sergeev, D.~E., {et~al.} 2021, arXiv:2109.11457
  [astro-ph, physics:physics].
\newblock \url{http://arxiv.org/abs/2109.11457}

\bibitem[{Venot {et~al.}(2018)Venot, Drummond, Miguel, Waldmann, Pascale, \&
  Zingales}]{venot_better_2018}
Venot, O., Drummond, B., Miguel, Y., {et~al.} 2018, Experimental Astronomy, 46,
  101, \dodoi{10.1007/s10686-018-9597-y}

\bibitem[{Villanueva {et~al.}(2018)Villanueva, Smith, Protopapa, Faggi, \&
  Mandell}]{villanueva_planetary_2018}
Villanueva, G.~L., Smith, M.~D., Protopapa, S., Faggi, S., \& Mandell, A.~M.
  2018, Journal of Quantitative Spectroscopy and Radiative Transfer, 217, 86,
  \dodoi{10.1016/j.jqsrt.2018.05.023}

\bibitem[{Walters {et~al.}(2019)Walters, Baran, Boutle, Brooks, Earnshaw,
  Edwards, Furtado, Hill, Lock, Manners, Morcrette, Mulcahy, Sanchez, Smith,
  Stratton, Tennant, Tomassini, Van~Weverberg, Vosper, Willett, Browse,
  Bushell, Carslaw, Dalvi, Essery, Gedney, Hardiman, Johnson, Johnson, Jones,
  Jones, Mann, Milton, Rumbold, Sellar, Ujiie, Whitall, Williams, \&
  Zerroukat}]{walters_met_2019}
Walters, D., Baran, A.~J., Boutle, I., {et~al.} 2019, Geoscientific Model
  Development, 12, 1909, \dodoi{10.5194/gmd-12-1909-2019}

\bibitem[{Wandel(2018)}]{wandel_biohabitability_2018}
Wandel, A. 2018, The Astrophysical Journal, 856, 165,
  \dodoi{10.3847/1538-4357/aaae6e}

\bibitem[{Wang {et~al.}(2020)Wang, Zhang, Hannachi, Hirooka, \&
  Hegglin}]{wang_tropical_2020}
Wang, T., Zhang, Q., Hannachi, A., Hirooka, T., \& Hegglin, M.~I. 2020,
  Quarterly Journal of the Royal Meteorological Society, 146, 2432,
  \dodoi{10.1002/qj.3801}

\bibitem[{Warner \& McIntyre(1996)}]{warner_propagation_1996}
Warner, C.~D., \& McIntyre, M.~E. 1996, Journal of the Atmospheric Sciences,
  53, 3213, \dodoi{10.1175/1520-0469(1996)053<3213:OTPADO>2.0.CO;2}

\bibitem[{Warner \& McIntyre(2001)}]{warner_ultrasimple_2001}
---. 2001, Journal of the Atmospheric Sciences, 58, 1837,
  \dodoi{10.1175/1520-0469(2001)058<1837:AUSPFN>2.0.CO;2}

\bibitem[{Watkins \& Cho(2010)}]{watkins_gravity_2010}
Watkins, C., \& Cho, J. Y.-K. 2010, The Astrophysical Journal, 714, 904,
  \dodoi{10.1088/0004-637X/714/1/904}

\bibitem[{Wolf(2017)}]{wolf_assessing_2017}
Wolf, E.~T. 2017, The Astrophysical Journal, 839, L1,
  \dodoi{10.3847/2041-8213/aa693a}

\bibitem[{Wordsworth(2015)}]{wordsworth_atmospheric_2015}
Wordsworth, R. 2015, The Astrophysical Journal, 806, 180,
  \dodoi{10.1088/0004-637X/806/2/180}

\bibitem[{Yang {et~al.}(2013)Yang, Cowan, \& Abbot}]{yang_stabilizing_2013}
Yang, J., Cowan, N.~B., \& Abbot, D.~S. 2013, The Astrophysical Journal, 771,
  L45, \dodoi{10.1088/2041-8205/771/2/L45}

\bibitem[{Yang {et~al.}(2020)Yang, Xie, \& Zhou}]{yang_occurrence_2020}
Yang, J.-Y., Xie, J.-W., \& Zhou, J.-L. 2020, The Astronomical Journal, 159,
  164, \dodoi{10.3847/1538-3881/ab7373}

\bibitem[{Yates {et~al.}(2020)Yates, Palmer, Manners, Boutle, Kohary, Mayne, \&
  Abraham}]{yates_ozone_2020}
Yates, J.~S., Palmer, P.~I., Manners, J., {et~al.} 2020, Monthly Notices of the
  Royal Astronomical Society, 492, 1691, \dodoi{10.1093/mnras/stz3520}

\bibitem[{Zhang {et~al.}(2000)Zhang, Koch, Davis, \&
  Kaplan}]{zhang_survey_2000}
Zhang, F., Koch, S.~E., Davis, C.~A., \& Kaplan, M.~L. 2000, Advances in
  Atmospheric Sciences, 17, 165, \dodoi{10.1007/s00376-000-0001-1}

\end{thebibliography}
\bibliographystyle{aasjournal}



\end{document}